\pdfoutput=1
\documentclass[11pt,a4paper]{article}
\usepackage[a4paper,left=2.4cm,right=2.4cm,top=2.5cm,bottom=2.4cm]{geometry}
\usepackage[T1]{fontenc}\usepackage[utf8]{inputenc}
\usepackage{mathpazo}                         
\usepackage[scaled=.92]{helvet}               

\usepackage{amsmath,amssymb,mathtools,bm}
\usepackage{microtype}
\usepackage{booktabs,array,longtable,colortbl,multirow,threeparttable}
\usepackage{graphicx,caption,float}
\usepackage[dvipsnames,table]{xcolor}
\usepackage{enumitem}
\usepackage{tikz}\usetikzlibrary{arrows.meta,positioning,shapes.geometric,fit,backgrounds,calc}
\usepackage[most]{tcolorbox}
\usepackage{titlesec,fancyhdr}
\usepackage[authoryear,round]{natbib}
\usepackage[colorlinks=true,linkcolor=linkblue,citecolor=linkblue,urlcolor=linkblue,pdfusetitle,bookmarksnumbered]{hyperref}
\usepackage[nameinlink,noabbrev]{cleveref}
\usepackage{needspace}\usepackage{setspace}
\definecolor{brand}{HTML}{16324F}
\definecolor{accent}{HTML}{D9531E}
\definecolor{linkblue}{HTML}{1F5CA8}
\definecolor{softblue}{HTML}{EAF1F8}
\definecolor{softorange}{HTML}{FDF0E8}
\definecolor{softgreen}{HTML}{E8F3EC}
\definecolor{rowshade}{HTML}{EEF2F6}
\definecolor{ink2}{HTML}{4A5563}
\titleformat{\section}{\sffamily\Large\bfseries\color{brand}}{\color{accent}\thesection}{.6em}{}[{\vspace{-.25em}\color{accent!55}\titlerule[.7pt]}]
\titleformat{\subsection}{\sffamily\large\bfseries\color{brand}}{\color{accent}\thesubsection}{.55em}{}
\titleformat{\subsubsection}{\sffamily\normalsize\bfseries\color{ink2}}{\thesubsubsection}{.5em}{}
\let\origsection\section\renewcommand{\section}{\needspace{9\baselineskip}\origsection}\let\origsubsection\subsection\renewcommand{\subsection}{\needspace{6\baselineskip}\origsubsection}
\titlespacing*{\section}{0pt}{1.6em}{.7em}\titlespacing*{\subsection}{0pt}{1.2em}{.4em}
\setlist{itemsep=2pt,topsep=4pt,leftmargin=1.6em}
\tcbset{enhanced jigsaw,breakable,boxrule=0pt,arc=2pt,left=7pt,right=7pt,top=5pt,bottom=5pt,fonttitle=\sffamily\bfseries}
\newtcbtheorem[number within=section]{proposition}{Proposition}{colback=softblue,colframe=brand,coltitle=white,colbacktitle=brand,attach boxed title to top left={yshift=-2mm,xshift=2mm},boxed title style={arc=2pt,boxrule=0pt},before skip=10pt,after skip=8pt,fonttitle=\sffamily\bfseries\small,breakable}{prop}
\newtcbtheorem[number within=section]{assumption}{Assumption}{colback=softorange,colframe=accent,coltitle=white,colbacktitle=accent,attach boxed title to top left={yshift=-2mm,xshift=2mm},boxed title style={arc=2pt,boxrule=0pt},before skip=8pt,after skip=6pt,fonttitle=\sffamily\bfseries\small}{ass}
\newtcolorbox{remarkbox}[1][]{colback=rowshade,colframe=ink2,leftrule=3pt,rightrule=0pt,toprule=0pt,bottomrule=0pt,arc=0pt,before skip=8pt,after skip=8pt,#1}
\newtcolorbox{keybox}[1][]{colback=softgreen,colframe=green!45!black,leftrule=3pt,rightrule=0pt,toprule=0pt,bottomrule=0pt,arc=0pt,before skip=8pt,after skip=8pt,#1}
\newtcolorbox{warnbox}[1][]{colback=softorange,colframe=accent,leftrule=3pt,rightrule=0pt,toprule=0pt,bottomrule=0pt,arc=0pt,before skip=8pt,after skip=8pt,#1}
\newenvironment{proofof}[1]{\par\noindent{\sffamily\bfseries\small\color{brand}Proof of #1.}\ }{\hfill$\square$\par\medskip}
\newcommand{\E}{\mathbb{E}}\newcommand{\Pp}{\mathbb{P}}\newcommand{\R}{\mathbb{R}}\newcommand{\Var}{\operatorname{Var}}
\newcommand{\expit}{\operatorname{expit}}\newcommand{\ATE}{\tau_0}

\newcommand{\code}[1]{\texttt{\small #1}}
\newcommand{\rmseOlsZero}{0.124}

\newcommand{\rmseClipZero}{0.092}

\newcommand{\covClipZero}{0.941}

\newcommand{\rmseOlsSix}{0.192}

\newcommand{\rmseTSix}{0.191}

\newcommand{\rmseTmixSix}{0.191}

\newcommand{\rmseClipSix}{0.107}

\newcommand{\rmseHuberSix}{0.191}

\newcommand{\rmseOlsTwenty}{0.292}

\newcommand{\rmseClipTwenty}{0.177}

\newcommand{\covClipTwenty}{0.919}
\newcommand{\seratioClipTwenty}{0.88}

\newcommand{\Breps}{1000}
\newcommand{\BrepsOther}{500}
\newcommand{\maxAbsBias}{0.013}
\newcommand{\mcseLo}{0.003}
\newcommand{\mcseHi}{0.009}
\newcommand{\maxRmseDiff}{0.002}
\newcommand{\covUncLo}{0.946}
\newcommand{\covUncHi}{0.961}
\newcommand{\clipRedLo}{26}
\newcommand{\clipRedHi}{45}

\newcommand{\artPosDiffOls}{+0.388}
\newcommand{\artCovPosDiffOls}{0.47}
\newcommand{\artPosDiffT}{+0.394}

\newcommand{\artPosDiffTmix}{+0.395}

\newcommand{\artPosDiffClip}{+0.173}
\newcommand{\artCovPosDiffClip}{0.61}
\newcommand{\artPosDiffHuber}{+0.394}

\newcommand{\artPosDiffIpw}{+0.384}
\newcommand{\artPosDiffGcomp}{+0.387}
\newcommand{\artOkMax}{0.013}
\newcommand{\misMaxAbs}{0.018}
\newcommand{\misRmseOutBad}{0.22}
\newcommand{\misRmseOk}{0.19}
\newcommand{\tailOneFiveOls}{1.003}
\newcommand{\tailOneFiveT}{0.989}
\newcommand{\tailOneFiveClip}{0.166}

\newcommand{\tailThreeOls}{0.163}
\newcommand{\tailThreeT}{0.162}

\newcommand{\tailThirtyDfOls}{0.093}

\newcommand{\tailThirtyDfClip}{0.093}

\newcommand{\tailSeRatioOneFive}{0.64}
\newcommand{\tailCovOneFive}{0.974}
\newcommand{\tailFactor}{6}
\newcommand{\slopeOls}{-0.53}
\newcommand{\slopeClip}{-0.52}
\newcommand{\sizeRatioLo}{0.56}
\newcommand{\sizeRatioHi}{0.59}
\newcommand{\sizeCovClipLo}{0.924}
\newcommand{\sizeCovClipHi}{0.946}
\newcommand{\asdZero}{0.119}
\newcommand{\sdOlsZero}{0.124}
\newcommand{\asdSix}{0.186}
\newcommand{\sdOlsSix}{0.192}
\newcommand{\asdTwelve}{0.235}
\newcommand{\sdOlsTwelve}{0.242}
\newcommand{\asdTwenty}{0.287}
\newcommand{\sdOlsTwenty}{0.291}
\newcommand{\emMinInc}{3.1\times10^{-7}}
\newcommand{\emNfits}{300}
\newcommand{\emNdec}{0}
\newcommand{\emIterMed}{65}
\newcommand{\emSigLowA}{0.008}
\newcommand{\emSigHighA}{15.2}
\newcommand{\emPiA}{0.51}
\newcommand{\emIterA}{25}
\newcommand{\emSigLowB}{22.4}
\newcommand{\emSigHighB}{22.4}

\newcommand{\emIterB}{18}
\newcommand{\emAppMinInc}{4.1\times10^{-7}}

\newcommand{\nSim}{1800}
\newcommand{\contSim}{5.6}
\newcommand{\plasTau}{5}
\newcommand{\Bplas}{500}
\newcommand{\plPoolBiasOls}{+0.05}
\newcommand{\plPoolRmseOls}{1.58}
\newcommand{\plPoolCovOls}{0.964}
\newcommand{\plPoolBiasT}{+0.15}

\newcommand{\plPoolCovT}{0.960}

\newcommand{\plPoolBiasClip}{+0.14}
\newcommand{\plPoolRmseClip}{1.55}

\newcommand{\plPoolBiasIpw}{-1.05}

\newcommand{\plArmBiasOls}{-0.10}
\newcommand{\plArmRmseOls}{2.34}

\newcommand{\plArmBiasT}{-0.15}

\newcommand{\plArmBiasClip}{+0.60}
\newcommand{\plArmRmseClip}{2.52}
\newcommand{\plArmCovClip}{0.950}

\newcommand{\plSkewZero}{1.28}
\newcommand{\plSkewOne}{-0.01}
\newcommand{\timeNmax}{16{,}000}
\newcommand{\timePipeMax}{0.37}
\newcommand{\timePipeSmall}{76}
\newcommand{\timeEcmMax}{74}
\newcommand{\timeOlsMax}{0.21}

\newcommand{\timeEcmRatio}{345}
\newcommand{\timeSlope}{0.72}
\newcommand{\machine}{x86\_64 / Python 3.13.18 / NumPy 2.5.3}
\newcommand{\nZonesAll}{519}
\newcommand{\nProvinces}{26}
\newcommand{\nApp}{1485}
\newcommand{\nZones}{362}
\newcommand{\nInt}{408}

\newcommand{\pctInt}{27.5}
\newcommand{\nReclass}{77}
\newcommand{\pctCeil}{79.6}
\newcommand{\pctHalf}{30.2}
\newcommand{\nOverHundred}{6}
\newcommand{\ratioMedPzq}{2.00}
\newcommand{\ratioMedInt}{1.00}

\newcommand{\pctRatioOneInt}{52}
\newcommand{\pctRatioTwoPzq}{37}
\newcommand{\pctRatioThreePzq}{17}
\newcommand{\naiveProg}{+18.8}

\newcommand{\appProgOls}{+17.4}
\newcommand{\appProgAbsOls}{17.4}
\newcommand{\appProgLoOls}{+14.2}
\newcommand{\appProgHiOls}{+21.3}
\newcommand{\appProgSeOls}{1.71}
\newcommand{\appProgSeIidOls}{1.88}

\newcommand{\appProgUncLo}{+17.3}
\newcommand{\appProgUncHi}{+17.8}
\newcommand{\splitProgLo}{+14.3}
\newcommand{\splitProgHi}{+19.7}
\newcommand{\trimProgLo}{+17.4}
\newcommand{\trimProgHi}{+17.4}
\newcommand{\kappaProgLo}{+17.9}
\newcommand{\kappaProgHi}{+20.7}

\newcommand{\appEpiOls}{+0.9}
\newcommand{\appEpiAbsOls}{0.9}
\newcommand{\appEpiLoOls}{+0.0}
\newcommand{\appEpiHiOls}{+3.0}
\newcommand{\appEpiSeOls}{0.87}
\newcommand{\appEpiSeIidOls}{0.72}

\newcommand{\appEpiUncLo}{+0.9}
\newcommand{\appEpiUncHi}{+1.5}
\newcommand{\splitEpiLo}{+0.6}
\newcommand{\splitEpiHi}{+2.4}

\newcommand{\appDenAbsOls}{24.5}
\newcommand{\appDenLoOls}{-30.2}
\newcommand{\appDenHiOls}{-16.1}

\newcommand{\appDenT}{-24.2}

\newcommand{\appDenClip}{-31.2}

\newcommand{\rawEpiClip}{+1.33}
\newcommand{\rawEpiSeIid}{0.10}

\newcommand{\rawEpiScale}{0.02}
\newcommand{\fbEpiScale}{4.2}

\newcommand{\denPctOls}{22}
\newcommand{\guardMadMc}{0}
\newcommand{\guardFloorMc}{108}

\newcommand{\guardMadApp}{27{,}835}
\newcommand{\guardFloorApp}{92{,}424}
\newcommand{\guardLstsqApp}{0}
\newcommand{\nSplits}{50}
\newcommand{\Bboot}{300}
\newcommand{\psMin}{0.05}
\newcommand{\psMax}{0.88}
\newcommand{\pctTrim}{1.8}
\newcommand{\smdMaxBefore}{0.54}
\newcommand{\smdMaxAfter}{0.12}
\newcommand{\nSmdOver}{3}
\newcommand{\appPctOut}{17.4}
\newcommand{\appKurt}{1.3}
\newcommand{\maxIfProgOls}{591}
\newcommand{\maxIfProgClip}{635}
\newcommand{\maxIfEpiOls}{585}
\newcommand{\maxIfEpiClip}{154}

\newcommand{\whoReqLast}{20.4}
\newcommand{\whoSacReqLast}{14.3}
\newcommand{\whoCovFirst}{11.9}
\newcommand{\whoCovPeak}{73.0}
\newcommand{\whoCovPeakYear}{2023}
\newcommand{\whoCovLast}{62.4}
\newcommand{\whoCovPrev}{38.5}

\newcommand{\afroMedLast}{43.1}
\newcommand{\nAfroLast}{31}
\newcommand{\owidYear}{2024}
\newcommand{\owidShare}{7.7}
\newcommand{\owidReq}{18.9}
\newcommand{\owidAfrica}{244}
\newcommand{\owidRank}{3}

\newcommand{\espSacTreatNineteen}{4.4}
\newcommand{\whoSacTreatNineteen}{7.6}

\hypersetup{pdftitle={Cross-fitted doubly robust estimation with Student-t outcome engines: ECM algorithms, Monte Carlo evidence and schistosomiasis programme data from the Democratic Republic of the Congo},pdfauthor={Cosmas Kambale Musambi; Ernest Fokou\'e},pdfsubject={stat.CO},pdfkeywords={computational statistics; doubly robust estimation; AIPW; cross-fitting; ECM algorithm; Student-t regression; plasmode simulation; schistosomiasis}}
\crefname{tcb@cnt@proposition}{Proposition}{Propositions}\crefname{tcb@cnt@assumption}{Assumption}{Assumptions}\crefname{figure}{Figure}{Figures}\crefname{table}{Table}{Tables}\crefname{section}{Section}{Sections}\crefname{proposition}{Proposition}{Propositions}\crefname{appendix}{Appendix}{Appendices}

\begin{document}
\begin{tcolorbox}[enhanced,colback=brand,colframe=brand,coltext=white,arc=3pt,boxrule=0pt,left=14pt,right=14pt,top=14pt,bottom=12pt,
  overlay={\fill[accent] (frame.south west) rectangle ([yshift=5pt]frame.south east);}]
{\sffamily\bfseries\LARGE\raggedright Cross-Fitted Doubly Robust Estimation with Student-$t$ Outcome Engines: ECM Algorithms, Monte Carlo Evidence and Schistosomiasis Programme Data from the Democratic Republic of the Congo\par}
\vspace{.8em}
{\sffamily\large Cosmas Kambale Musambi$^{1}$\quad and\quad Ernest Fokou\'e$^{2}$\par}
\vspace{.3em}
{\sffamily\small $^{1}$Universit\'e de Dschang, Data Science and Mathematical Engineering (DSEM) \quad\href{mailto:ckambale86@gmail.com}{\textcolor{white}{\underline{ckambale86@gmail.com}}}\\
$^{2}$School of Mathematics and Statistics, Rochester Institute of Technology \quad\href{mailto:ernest.fokoue@rit.edu}{\textcolor{white}{\underline{ernest.fokoue@rit.edu}}}\par}
\vspace{.2em}{\sffamily\small 13 November 2026\par}
\end{tcolorbox}
\vspace{.4em}
\begin{tcolorbox}[colback=softblue,colframe=brand,leftrule=3pt,rightrule=0pt,toprule=0pt,bottomrule=0pt,arc=0pt]
\noindent{\sffamily\bfseries\color{brand} Abstract.}\ Routine data on preventive chemotherapy for schistosomiasis are confounded, clustered, heavy-tailed and heaped. We study how cross-fitted augmented inverse probability weighted (AIPW) estimators behave, and how they should be computed, when least squares is replaced by Student-$t$ outcome engines: Huber and $t$ iteratively reweighted least squares, and a two-component $t$ scale-mixture regression fitted by an ECM algorithm, alongside a residual-clipped augmentation. We give four short propositions (exact AIPW bias, the efficiency bound, the estimand of clipping, monotone ascent of a scale-constrained ECM), a single-file Python implementation with cluster-level cross-fitting, cluster-robust standard errors and a cluster bootstrap, and a Monte Carlo study with up to \Breps{} replications per cell. Robust outcome engines leave AIPW unchanged (RMSE differences at most \maxRmseDiff), whereas clipping cuts RMSE by \clipRedLo--\clipRedHi\% under symmetric contamination, under-covers, and, as a plasmode experiment built on real covariates and residuals shows, shifts the estimand when residual laws differ between arms. One-sided artefacts that differ between arms bias every estimator. Heaped data also break naive code: the MAD scale collapses and the mixture likelihood is unbounded, and we give simple guards. Applied to \nApp{} rounds in \nZones{} health zones of the Democratic Republic of the Congo (ESPEN, 2014--2025), integrated delivery appears to raise programme coverage by \appProgAbsOls{} percentage points, yet epidemiological coverage differs by only \appEpiAbsOls: the gap comes from targeting denominators, a real differential artefact that no outcome engine can remove.

\smallskip\noindent{\sffamily\bfseries\color{brand} Keywords:} computational statistics; doubly robust estimation; AIPW; cross-fitting; ECM algorithm; Student-$t$ regression; plasmode simulation; cluster bootstrap; heaped data; schistosomiasis; Democratic Republic of the Congo.\\
{\sffamily\bfseries\color{brand} MSC 2020:} 62-08 (primary); 65C05, 62D20, 62F35, 62P10. \quad {\sffamily\bfseries\color{brand} arXiv subject class:} stat.CO (cross-lists stat.ME, stat.AP).
\end{tcolorbox}
\begin{warnbox}[title=\sffamily\bfseries\small Data,fonttitle=\sffamily\bfseries\small,coltitle=accent,colbacktitle=softorange,attach title to upper={\ }]
\small The empirical analyses use public programme records: the ESPEN implementation-unit file for schistosomiasis in the Democratic Republic of the Congo (DRC; \nZonesAll{} health zones, 2014--2025), the WHO preventive chemotherapy databank and two WHO series redistributed by Our World in Data \citep{espen,who_pct,owid_req,owid_cov}. These are aggregate counts per health zone and year; no record concerns an individual. The Monte Carlo experiments draw from a stated simulation model, and the plasmode experiment reuses the real DRC covariates and residuals with an effect fixed by design.
\end{warnbox}

\section{Introduction}\label{sec:intro}
Schistosomiasis remains one of the heaviest neglected tropical diseases (NTDs) in sub-Saharan Africa \citep{colley2014}, and the Democratic Republic of the Congo carries a large share of the continent's need for preventive chemotherapy (PC) with praziquantel. WHO recommends annual PC wherever prevalence reaches 10\% and sets a coverage target of at least 75\% of those requiring treatment \citep{who2022}. Programmes report what they deliver to WHO and to the Expanded Special Project for Elimination of Neglected Tropical Diseases (ESPEN), which publishes the counts health zone by health zone \citep{who_pct,espen}. Anyone who tries to learn from these records quickly meets three problems at once: delivery choices are confounded by endemicity, geography and history; outcomes are heaped, capped and occasionally impossible (coverage above 100\%); and the denominators behind a percentage are themselves planning decisions.

The augmented inverse probability weighted (AIPW) estimator is the usual tool for confounded comparisons of this kind. It combines an outcome regression with a propensity model and stays consistent if either one is right \citep{robins1994,bang2005}; with cross-fitting, flexible nuisance fits keep valid inference \citep{chernozhukov2018}. Because programme outcomes are heavy-tailed, it is tempting to replace least squares by a Student-$t$ regression \citep{lange1989} or by a mixture of $t$ components \citep{peel2000}. This paper asks a computational question: \emph{what does such a substitution actually change in the estimator, at what cost, and how should it be implemented when the data are clustered, heaped and capped?}

\paragraph{Contributions.} The paper is written for statistical computing and makes five contributions.
\begin{enumerate}[label=\textbf{(C\arabic*)}]
\item \textbf{An implementation.} Cross-fitted AIPW with four interchangeable outcome engines (OLS, Huber IRLS, fixed-$\nu$ Student-$t$ IRLS and a two-component Student-$t$ scale-mixture regression fitted by an ECM algorithm), a residual-clipped variant, cluster-level cross-fitting, cluster-robust standard errors and a cluster bootstrap. Everything sits in one Python file (\cref{app:repro}).
\item \textbf{Four propositions with proofs} (\cref{sec:theory}): the exact bias of AIPW under misspecification, why a robust nuisance fit cannot beat the efficiency bound of the mean-ATE problem, the estimand implied by clipping, and monotone ascent of the (constrained) ECM algorithm. These are known or elementary results; we restate them because the computations are read through them, and we claim no new asymptotic theorem.
\item \textbf{Monte Carlo evidence} (\cref{sec:design,sec:mc}): \Breps{} replications per cell for the main experiment and \BrepsOther{} elsewhere, with Monte Carlo standard errors \citep{morris2019}; a plasmode experiment that keeps the real DRC covariates and residual laws \citep{franklin2014}; ECM diagnostics; and a computing-time benchmark.
\item \textbf{Numerical lessons from heaped data} (\cref{sec:numerics}): the MAD scale collapses when a robust fit passes through a heap, and the mixture likelihood becomes unbounded when outcomes are tied. Both failures occur in the DRC data, and both have simple remedies.
\item \textbf{An application} to \nApp{} rounds of PC for school-age children in \nZones{} health zones (\cref{sec:app}). The apparent advantage of integrated delivery (praziquantel with albendazole or mebendazole) on reported programme coverage turns out to be almost entirely a difference in targeting denominators, a real instance of the differential-artefact problem that our simulations show no outcome engine can repair.
\end{enumerate}

\Cref{sec:data} describes the data, \cref{sec:framework} the causal set-up, \cref{sec:methods,sec:numerics} the estimators and their computation, \cref{sec:theory} the theory, \cref{sec:design,sec:mc} the Monte Carlo study, \cref{sec:app} the application and \cref{sec:drc} a target-trial template for future work. \Cref{sec:discussion} draws the practical conclusions.

\section{Schistosomiasis programme data}\label{sec:data}
\subsection{Sources}
We use four public files, all downloaded on 13 November 2026 and distributed with the code. (i)~The \textbf{ESPEN implementation-unit file} for schistosomiasis in the DRC \citep{espen} gives, for each of \nZonesAll{} health zones in \nProvinces{} provinces and each year 2014--2025, the endemicity class, the populations requiring PC, targeted and treated, the derived coverages and the drug package, separately for pre-school children, school-age children (SAC), adults and the total. (ii)~The \textbf{WHO PC databank} \citep{who_pct} gives the same indicators at country level for 59 countries, 2006--2025. (iii--iv)~\textbf{Our World in Data} redistributes two WHO series, the population requiring PC and national coverage \citep{owid_req,owid_cov}. A second ESPEN extract covering 2020--2025 was also supplied to us; it is a strict subset of file (i), row for row, and is not used separately.

\Cref{tab:data} lists what each source can and cannot support. None of them contains infection outcomes, so none can identify an effect of treatment on disease. What they do support is the study of programme delivery, and that is where the statistical difficulties of this paper come from.

\begin{table}[!htb]\centering\small
\caption{Data used in this paper and what each source can support.}\label{tab:data}
\begin{tabular}{@{}>{\raggedright\arraybackslash}p{3.7cm}>{\raggedright\arraybackslash}p{5.2cm}>{\raggedright\arraybackslash}p{5.5cm}@{}}
\toprule \rowcolor{rowshade}\textbf{Object}&\textbf{Content}&\textbf{Use in this paper}\\\midrule
ESPEN health-zone file, DRC&\nZonesAll{} zones $\times$ 12 years $\times$ 4 target groups; counts, coverage, endemicity, drug package&Descriptive analysis; application (\cref{sec:app}); plasmode design (E6)\\
WHO PC databank&Country-year indicators, 2006--2025&National trends; consistency check against ESPEN\\
Our World in Data (WHO series)&Population requiring PC; national coverage&The DRC in the African context\\
Simulation model&Stated in \cref{sec:design}; known ATE $=0.35$&Monte Carlo experiments E1--E5, E7, E8\\
Infection or morbidity outcomes&\emph{Not available}&No effect on disease is estimated\\\bottomrule
\end{tabular}\end{table}

\subsection{The national picture}
\begin{figure}[!htb]\centering\includegraphics[width=\textwidth]{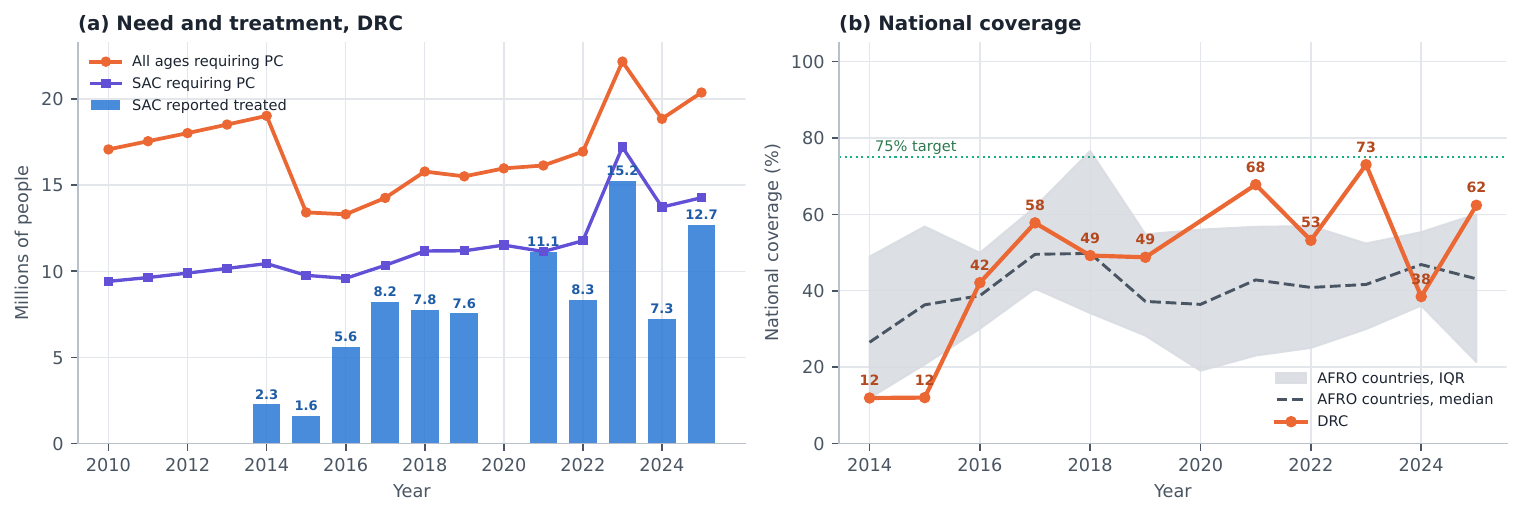}
\caption{WHO national series for the DRC. (a) Population requiring PC (all ages and SAC, lines) and SAC reported treated (bars, millions, labelled). (b) National coverage in the DRC (orange, labelled) against the median and interquartile range of countries in the WHO African Region; the dotted line marks the 75\% target. Source: WHO PC databank \citep{who_pct}.}\label{fig:who}\end{figure}

\begin{figure}[!htb]\centering\includegraphics[width=\textwidth]{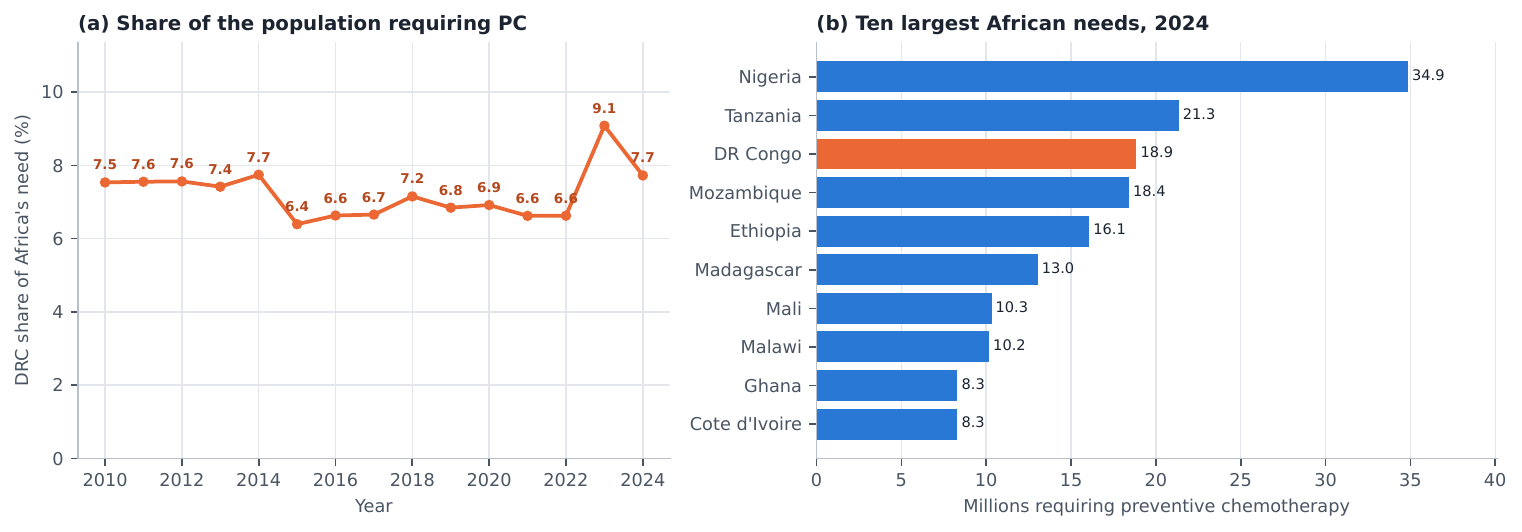}
\caption{The DRC in the African context. (a) The DRC's share of the African population requiring PC for schistosomiasis, by year. (b) The ten African countries with the largest population requiring PC in \owidYear{} (millions); the DRC is in orange. Source: Our World in Data, WHO series \citep{owid_req}.}\label{fig:africa}\end{figure}

In 2025 WHO counted \whoReqLast{} million people in the DRC requiring PC for schistosomiasis, \whoSacReqLast{} million of them school-age children (\cref{fig:who}a). National coverage rose from \whoCovFirst\% in 2014 to a peak of \whoCovPeak\% in \whoCovPeakYear, fell to \whoCovPrev\% in 2024 and recovered to \whoCovLast\% in 2025, against a median of \afroMedLast\% across the \nAfroLast{} countries of the WHO African Region that reported in 2025 (\cref{fig:who}b). No SAC round was reported in 2020, the first year of the COVID-19 pandemic. On the Our World in Data series the DRC accounted for \owidShare\% of Africa's population requiring PC in \owidYear{} (\owidReq{} of \owidAfrica{} million), the \owidRank{}th largest need on the continent (\cref{fig:africa}).


\subsection{What the health-zone file looks like}
\begin{figure}[!htb]\centering\includegraphics[width=\textwidth]{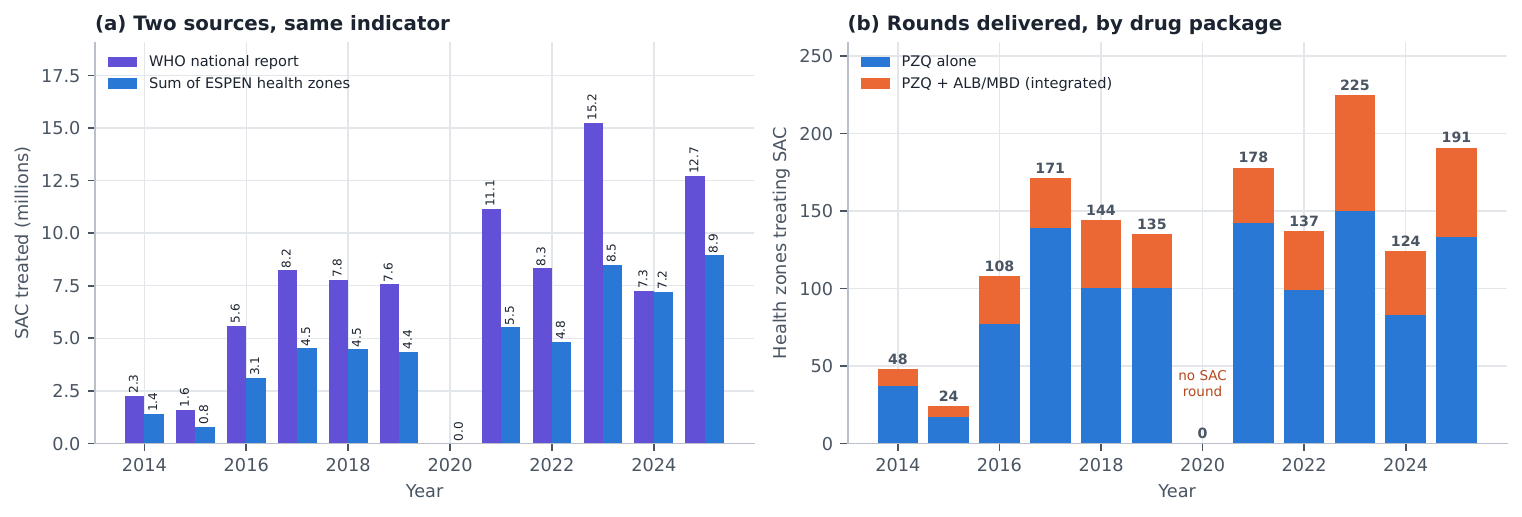}
\caption{Two sources, one indicator. (a) SAC treated according to the WHO national report and to the sum of ESPEN health-zone records (millions, labelled). (b) Health zones that delivered a SAC round, split by drug package; totals above the bars. No SAC round was recorded in 2020.}\label{fig:espen}\end{figure}

\begin{figure}[!htb]\centering\includegraphics[width=.92\textwidth]{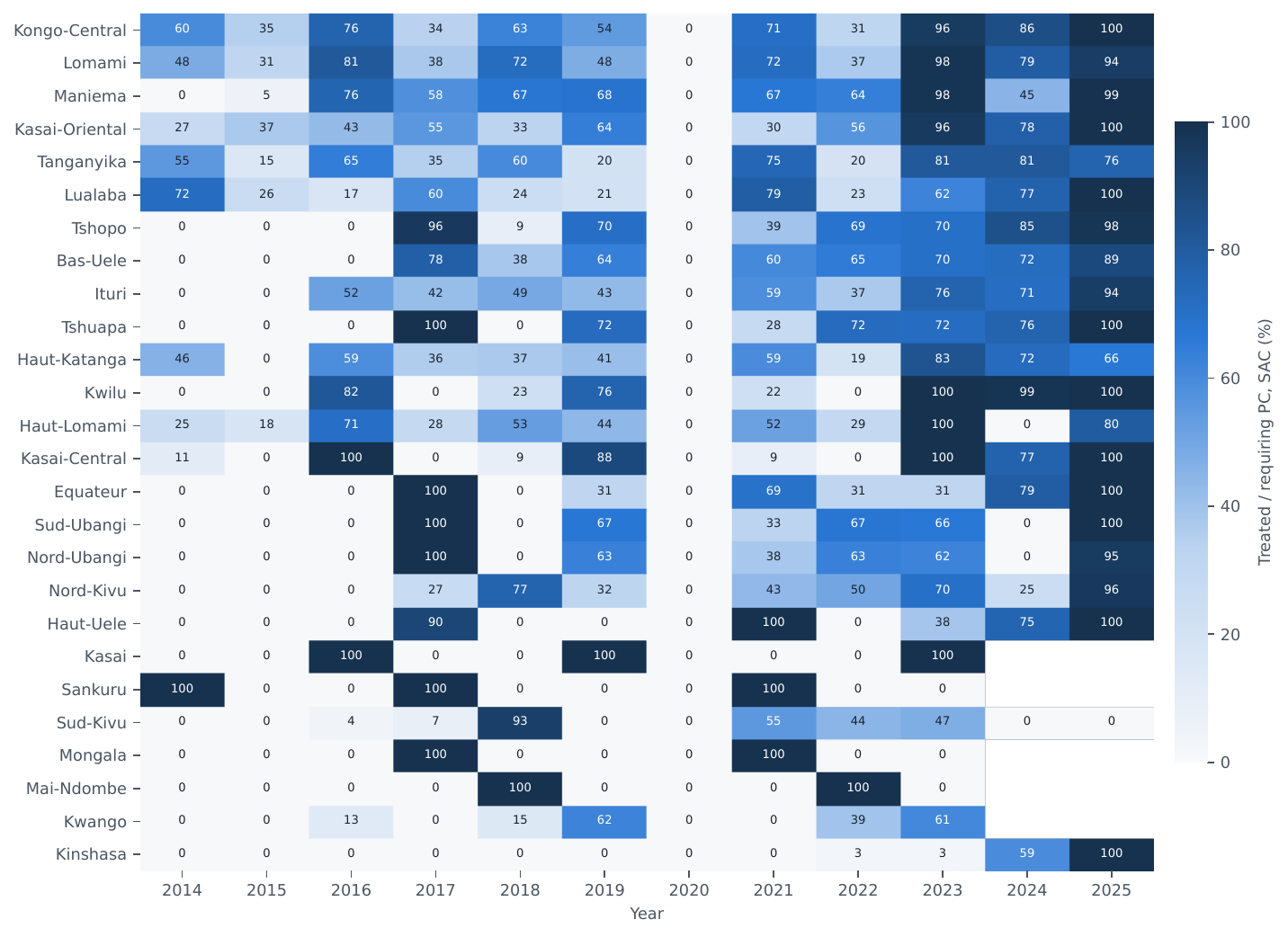}
\caption{Epidemiological coverage of school-age children by province and year: SAC treated divided by SAC requiring PC, summed over the province's health zones (\%). Provinces are ordered by their mean; blank cells have no SAC requiring PC. Source: ESPEN \citep{espen}.}\label{fig:heatmap}\end{figure}

\begin{figure}[!htb]\centering\includegraphics[width=\textwidth]{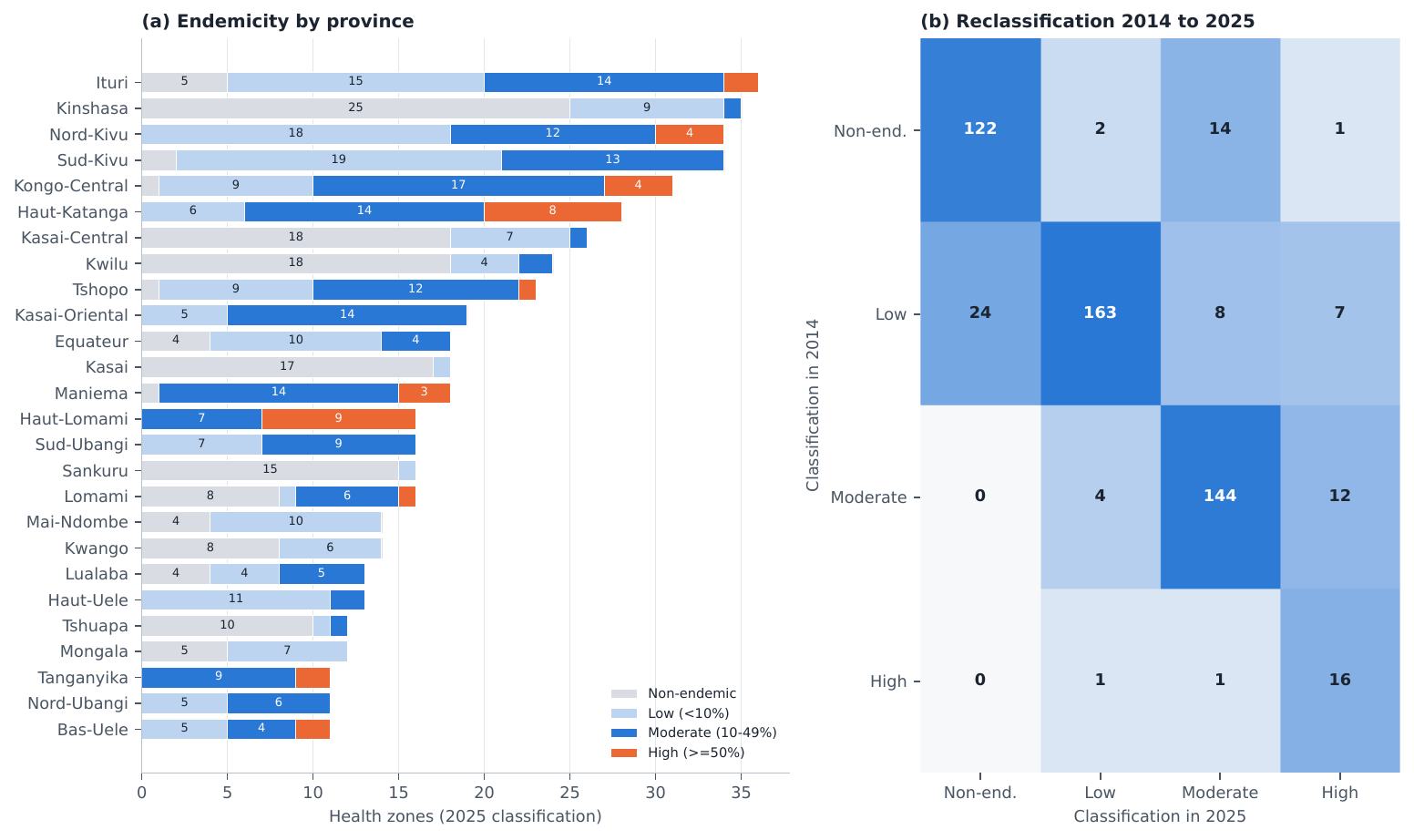}
\caption{Endemicity of the \nZonesAll{} health zones. (a) Classification in 2025 by province (counts labelled). (b) Classification in 2014 against 2025; off-diagonal cells are reclassified zones.}\label{fig:endemic}\end{figure}

Health-zone records do not simply add up to the national report. For 2019, WHO lists \whoSacTreatNineteen{} million SAC treated, whereas the ESPEN zones sum to \espSacTreatNineteen{} million (\cref{fig:espen}a). The gap varies from year to year and is a first reminder that the two files are produced by different reporting chains. \Cref{fig:espen}b shows how many zones delivered a SAC round each year and with which drug package. Coverage differs strongly between provinces and years (\cref{fig:heatmap}), and endemicity classes are not static: \nReclass{} zones changed class at least once between 2014 and 2025 (\cref{fig:endemic}).


\begin{figure}[!htb]\centering\includegraphics[width=\textwidth]{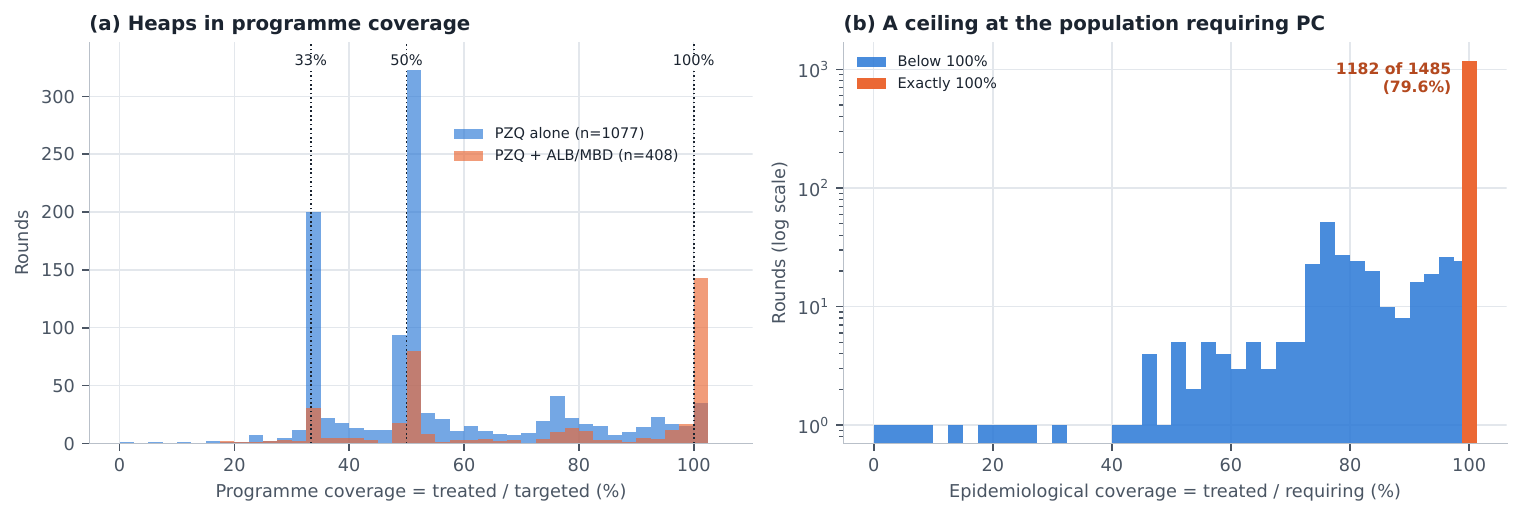}
\caption{Heaping and ceilings in the \nApp{} delivered SAC rounds. (a) Programme coverage by drug package; dotted lines mark the heaps at 33\%, 50\% and 100\%. (b) Epidemiological coverage on a log count scale; the orange bar counts rounds at exactly 100\%.}\label{fig:heaping}\end{figure}

Two features of the file matter directly for estimation (\cref{fig:heaping}). First, the reported number treated never exceeds the population requiring PC and equals it exactly in \pctCeil\% of delivered rounds, so epidemiological coverage piles up at 100\%. Second, programme coverage (treated over targeted) is heaped: \pctHalf\% of rounds sit exactly at 50\%, others at 33\% or 100\%, and \nOverHundred{} rounds exceed 100\%. These are coarsened data in the sense of \citet{heitjan1991}, and they echo long-standing concerns about reported MDA coverage \citep{worrell2012}. For a regression engine they produce residual distributions with a sharp spike and long tails, the setting in which robust and $t$-based fits are usually recommended.

\begin{figure}[!htb]\centering\includegraphics[width=\textwidth]{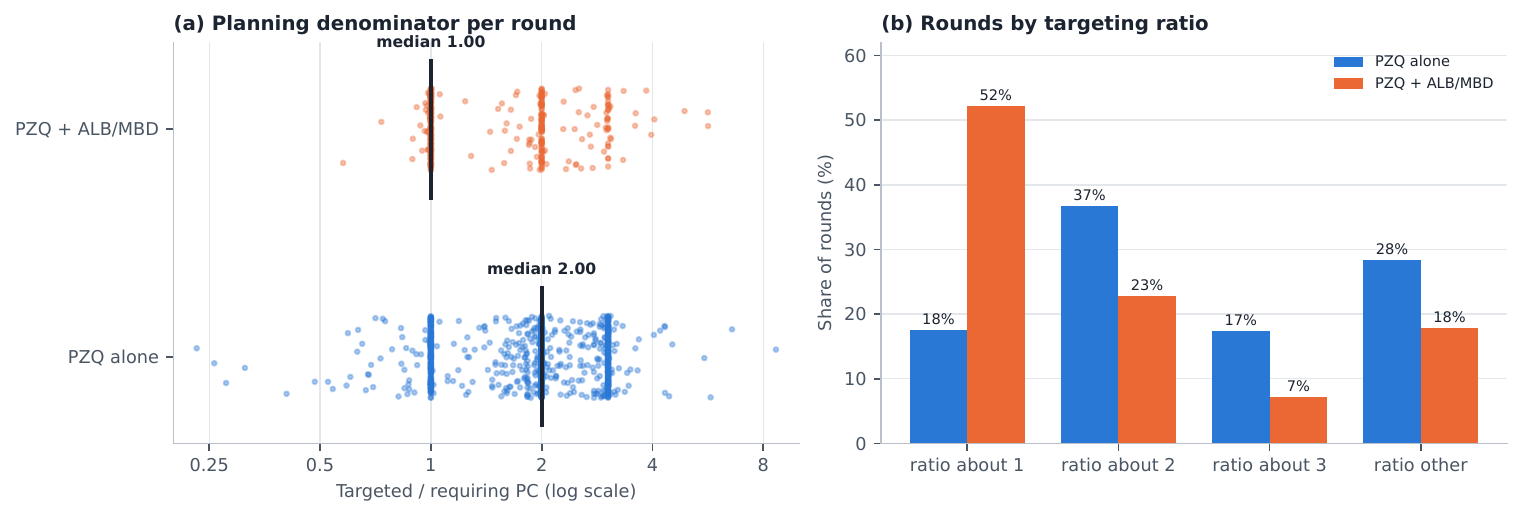}
\caption{The targeting denominator differs by drug package. (a) Ratio of the population targeted to the population requiring PC in each round (log scale, jittered), with medians. (b) Share of rounds whose ratio lies within 2\% of 1, 2 or 3, or elsewhere.}\label{fig:denominator}\end{figure}

The third feature is the one that drives the application. The planning denominator, the number targeted, is not a fixed multiple of need. In rounds that used praziquantel alone the median ratio of targeted to requiring is \ratioMedPzq{} (\pctRatioTwoPzq\% of these rounds have a ratio within 2\% of two and \pctRatioThreePzq\% within 2\% of three), whereas in integrated rounds it is \ratioMedInt{}, with \pctRatioOneInt\% of rounds targeting exactly the population in need (\cref{fig:denominator}). Because programme coverage divides by the targeted number, the same treated count yields a much lower percentage under the first convention than under the second.

\section{Causal framework}\label{sec:framework}
For unit $i$ let $A_i\in\{0,1\}$ be a well-defined intervention, $X_i\in\R^p$ pre-treatment confounders, $Y_i$ the outcome and $Y_i(a)$ the potential outcome under $A_i=a$. The target is the average treatment effect (ATE)
\begin{equation}\ATE=\E\{Y(1)-Y(0)\}.\label{eq:ate}\end{equation}
Write $m_a(x)=\E(Y\mid A=a,X=x)$ and $e(x)=\Pp(A=1\mid X=x)$.
\begin{assumption}{Consistency and well-defined treatment}{consistency}$Y=Y(A)$, and the versions of each treatment are specified precisely enough for this to hold.\end{assumption}
\begin{assumption}{Conditional exchangeability}{exch}$\{Y(0),Y(1)\}\perp A\mid X$.\end{assumption}
\begin{assumption}{Positivity}{pos}There is $\epsilon>0$ with $\epsilon<e(X)<1-\epsilon$ almost surely.\end{assumption}
\begin{assumption}{No interference}{sutva}A unit's outcome does not depend on other units' treatments, or interference is built into the estimand.\end{assumption}
Under \cref{ass:consistency,ass:exch,ass:pos,ass:sutva} and $\E|Y|<\infty$, $\ATE=\E\{m_1(X)-m_0(X)\}$ \citep{rubin1974,rosenbaum1983,hernan2020}. None of these assumptions can be checked from the observed data alone. Community-wide PC changes transmission, so \cref{ass:sutva} is implausible for individuals; in the application the unit is a health-zone round and the outcome a delivery indicator, for which spill-overs are less of a concern \citep{hudgens2008}. Repeated rounds in the same zone are dependent, which we handle by clustering (\cref{sec:clusters}). Post-treatment variables must never enter $X$.

\section{Estimators}\label{sec:methods}
\subsection{Cross-fitted AIPW}
Split the sample into $K$ folds. For observation $i$ in fold $k(i)$, fit the nuisance functions $\hat\eta^{(-k)}=(\hat m_0,\hat m_1,\hat e)$ on the other folds and evaluate
\begin{align}
\psi(W_i;\hat\eta)&=\hat m_1(X_i)-\hat m_0(X_i)+\frac{A_i}{\hat e(X_i)}\{Y_i-\hat m_1(X_i)\}-\frac{1-A_i}{1-\hat e(X_i)}\{Y_i-\hat m_0(X_i)\},\label{eq:aipw}\\
\hat\tau&=\frac1n\sum_{i=1}^n\psi(W_i;\hat\eta^{(-k(i))}),\qquad \widehat{\mathrm{SE}}=\frac{\operatorname{sd}\{\psi_i\}}{\sqrt n}.\label{eq:se}
\end{align}
The propensity is a cross-fitted logistic regression truncated to $[0.05,0.95]$, a numerical safeguard that may shift the target population \citep{crump2009}. The simulations use $K=3$ i.i.d.\ folds; the application uses $K=5$ folds formed at the cluster level.

\begin{figure}[!htb]\centering
\begin{tikzpicture}[font=\sffamily\scriptsize,node distance=7mm and 6mm,
  box/.style={draw=brand,fill=softblue,rounded corners=2pt,align=center,minimum height=9mm,minimum width=24mm,line width=.6pt},
  hi/.style={draw=accent,fill=softorange,rounded corners=2pt,align=center,minimum height=9mm,minimum width=24mm,line width=.6pt},
  ok/.style={draw=green!45!black,fill=softgreen,rounded corners=2pt,align=center,minimum height=9mm,minimum width=24mm,line width=.6pt},
  arr/.style={-{Stealth[length=2mm]},line width=.7pt,color=brand}]
\node[box] (data) {Data\\$(X_i,A_i,Y_i)$,\\cluster $g_i$};
\node[box,right=of data] (split) {Split into $K$ folds\\(whole clusters)};
\node[box,right=of split,yshift=10mm] (prop) {Propensity $\hat e$\\logistic, truncated};
\node[hi,right=of split,yshift=-10mm] (out) {Outcome $\hat m_a$ per arm\\OLS | Huber | $t$-IRLS | $t$-mixture ECM};
\node[box,right=10mm of prop,yshift=-10mm,xshift=8mm] (score) {Held-out fold:\\score $\psi_i$ in \eqref{eq:aipw}\\(optional clip)};
\node[ok,right=of score] (est) {$\hat\tau=\bar\psi$\\cluster SE,\\cluster bootstrap};
\draw[arr] (data)--(split);\draw[arr] (split)--(prop);\draw[arr] (split)--(out);
\draw[arr] (prop)--(score);\draw[arr] (out)--(score);\draw[arr] (score)--(est);
\node[font=\sffamily\tiny\itshape,color=ink2,above=1mm of prop] {fit on training folds};
\node[font=\sffamily\tiny\itshape,color=ink2,below=1mm of out] {fit on training folds};
\end{tikzpicture}
\caption{Estimation workflow. Nuisance functions are fitted on training folds and evaluated on the held-out fold, so no observation is scored by a model that has seen its own outcome. With clustered data, whole clusters move between folds together. The outcome engine (orange) is the only component that differs between the AIPW variants compared here.}\label{fig:workflow}
\end{figure}
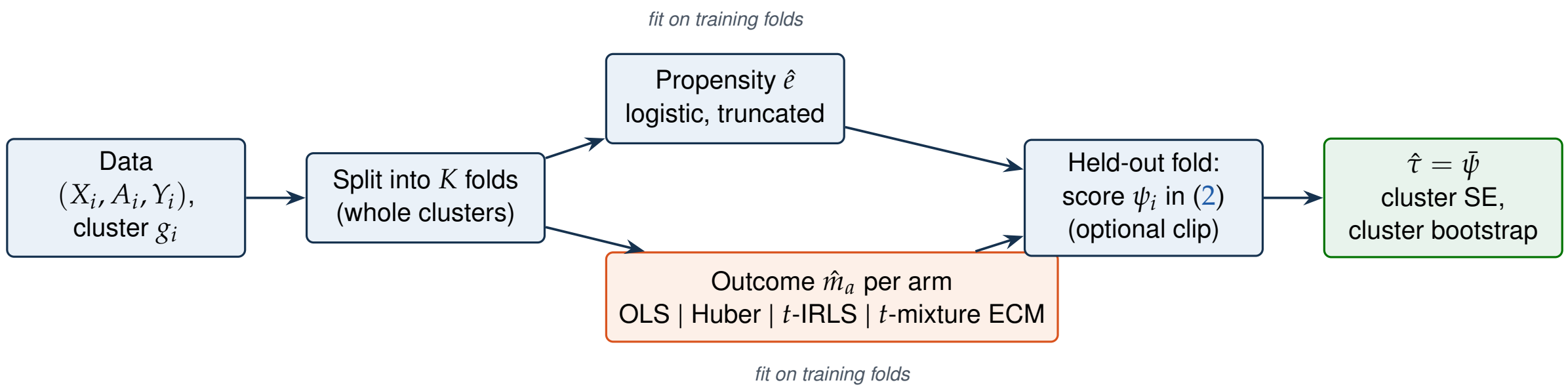

\subsection{Outcome engines}
Within each arm let $D_i=(1,X_i^\top)^\top$ and $Y_i=\beta^\top D_i+\varepsilon_i$.
\paragraph{OLS.} Ordinary least squares, the default choice.
\paragraph{Huber IRLS.} $M$-estimation with Huber's $\psi$ \citep{huber1964}, tuning constant $c=1.345$ and MAD scale, computed by iteratively reweighted least squares \citep{hampel1986}.
\paragraph{Student-$t$ IRLS (fixed $\nu$).} Working likelihood $\varepsilon_i\sim t_\nu(0,\sigma^2)$ with $\nu=4$ and update
\begin{equation}w_i^{(r)}=\frac{\nu+1}{\nu+\{(Y_i-\beta^{(r)\top}D_i)/\sigma^{(r)}\}^2},\qquad\beta^{(r+1)}=(D^\top W^{(r)}D+\lambda I)^{-1}D^\top W^{(r)}Y,\label{eq:irls}\end{equation}
with a stabilizing ridge $\lambda=10^{-5}$ and a weighted-residual scale update. Large residuals receive small weights.
\paragraph{Student-$t$ scale-mixture regression (ECM).} A latent class $Z_i\in\{1,\dots,K\}$ ($K=2$) and a latent precision $U_i$ generate
\begin{equation}Z_i\sim\mathrm{Cat}(\pi),\qquad U_i\mid Z_i=k\sim\mathrm{Gamma}\!\big(\tfrac{\nu}{2},\tfrac{\nu}{2}\big),\qquad Y_i\mid U_i,Z_i=k\sim N\!\big(\beta^\top D_i,\ \sigma_k^2/U_i\big),\label{eq:mix}\end{equation}
so that marginally $Y_i\mid Z_i=k\sim t_\nu(\beta^\top D_i,\sigma_k^2)$ \citep{peel2000}. The coefficients $\beta$ are shared and the components differ in scale; $\nu=4$ is fixed. The E-step is exact:
\begin{equation}r_{ik}=\Pp(Z_i=k\mid y_i),\qquad u_{ik}=\E(U_i\mid y_i,Z_i=k)=\frac{\nu+1}{\nu+\{(y_i-\beta^\top D_i)/\sigma_k\}^2}.\label{eq:estep}\end{equation}
Two conditional-maximization steps follow \citep{meng1993}: $\beta=(D^\top WD)^{-1}D^\top Wy$ with $w_i=\sum_k r_{ik}u_{ik}/\sigma_k^2$, then $\sigma_k^2=\max\{\sum_i r_{ik}u_{ik}(y_i-\beta^\top D_i)^2/\sum_i r_{ik},\ \underline{\sigma}^2\}$ and $\pi_k=n^{-1}\sum_i r_{ik}$. The lower bound $\underline\sigma=10^{-3}s_0$, with $s_0$ the MAD of the OLS residuals, is explained in \cref{sec:numerics}. Because the posterior of $(Z,U)$ is available in closed form, EM is the special case of variational inference in which the variational family contains the true posterior \citep{blei2017}; a variational mixture of experts with covariate-dependent gating \citep{jordan1994} is \emph{not} implemented (\cref{app:vb}).

\subsection{Clipped augmentation and baselines}
To bound the influence of single residuals we also study \textbf{AIPW-$t$clip}, which replaces $Y-\hat m_a(X)$ in \eqref{eq:aipw} by $\mathrm{clip}\{Y-\hat m_a(X),\pm\kappa\hat s_a\}$ with $\kappa=3$ fixed in advance and $\hat s_a$ the MAD scale of the training-fold residuals of the $t$ fit. This is a plain winsorization of the augmentation term, not a new method, and it changes the estimand unless the residuals are symmetric (\cref{prop:clip}). Two baselines that are not doubly robust complete the comparison: the \textbf{G-formula} $n^{-1}\sum_i\{\hat m_1(X_i)-\hat m_0(X_i)\}$ with OLS, and the H\'ajek \textbf{IPW} estimator.

\subsection{Clustered data}\label{sec:clusters}
In the application the unit is a health-zone round and the same zone contributes up to eleven rounds. We therefore (i) assign whole zones to folds, so that no zone is used to fit the nuisance functions that score its own rounds; (ii) report the cluster-robust standard error
\begin{equation}\widehat{\mathrm{SE}}_{\rm cl}=\frac1n\Big\{\frac{G}{G-1}\sum_{g=1}^G\Big(\sum_{i\in g}(\psi_i-\hat\tau)\Big)^2\Big\}^{1/2}\label{eq:secl}\end{equation}
over the $G$ zones \citep{liang1986,cameron2015}, which is the one-way version of the variance of \citet{chiang2022}; and (iii) compute percentile intervals from a cluster bootstrap that resamples zones with replacement and repeats the \emph{whole} cross-fitting pipeline in every resample, with duplicated zones treated as distinct clusters.

\section{Computation}\label{sec:numerics}
\subsection{Cost}
Each engine solves $p\times p$ weighted least-squares systems, so one iteration costs $O(np^2+p^3)$. OLS needs one such solve, the IRLS engines a few dozen, and the ECM algorithm adds an E-step of cost $O(nK)$ per iteration. A full cross-fitted run fits one propensity model and $4\times2$ outcome models in each of the $K$ folds and is therefore linear in $n$ for fixed $p$; experiment E8 (\cref{sec:timing}) measures the constants.

\subsection{Two failures on heaped data, and their remedies}
The DRC outcomes exposed two numerical problems that do not arise with continuous simulated data.

\paragraph{Collapse of the MAD.} When more than half of the outcomes in an arm sit on the same value, as with epidemiological coverage at 100\%, a robust fit tends to pass through the heap. More than half of its residuals are then essentially zero, the MAD collapses, and anything scaled by it breaks down. For the clipped estimator the consequence is silent: every residual is clipped to almost nothing, the augmentation term vanishes, and AIPW-$t$clip quietly turns into a plug-in regression estimator. On the epidemiological-coverage outcome the median clipping scale computed from the raw MAD was \rawEpiScale{} percentage points, the clipped estimate was \rawEpiClip{} and its i.i.d.\ standard error was \rawEpiSeIid, far smaller than that of any unclipped estimator and a misleading sign of precision. For continuous data the ratio of the normalized MAD to the mean absolute deviation is of order one (between 0.4 and 1.4 for programme coverage and the targeting ratio in the DRC data); for epidemiological coverage it fell below 0.03. Our implementation therefore treats a ratio below 0.05 as a collapse and switches to the normalized mean absolute deviation, which gives a median clipping scale of \fbEpiScale{} points. The rule was triggered in \guardMadMc{} scale evaluations over the whole Monte Carlo study, so the simulation results are unaffected by it, and in \guardMadApp{} evaluations across the DRC analyses (main fits, bootstrap and sensitivity analyses), all of them on the epidemiological-coverage outcome.

\paragraph{Unbounded mixture likelihood.} The scale-mixture likelihood is unbounded: a component centred on tied observations can shrink its scale towards zero while its likelihood grows without limit. With heaped outcomes this is what the ECM iterations do, and in one cluster-bootstrap resample the weighted least-squares system became singular. We bound the component scales below by $\underline\sigma=10^{-3}s_0$ and fall back to least squares if a linear system is singular; the bound was active in \guardFloorMc{} ECM iterations of the Monte Carlo study (all in heavy-tailed cells, where it changed no reported estimate by more than $10^{-5}$) and in \guardFloorApp{} iterations of the DRC analyses, and once it was in place the least-squares fallback was needed \guardLstsqApp{} times. The constrained CM-step for $\sigma_k^2$ is still an exact maximization (the expected complete-data log-likelihood is unimodal in $\sigma_k^2$, so its constrained maximizer is the clipped unconstrained one), and the ascent property of \cref{prop:em} is preserved. \Cref{sec:mc_em} shows what the bound does on the DRC data.

\section{Theoretical properties}\label{sec:theory}
The results below are standard or elementary. We state them precisely because the computations are interpreted through them; proofs are in \cref{app:proofs}.

\begin{proposition}{Exact bias of AIPW under misspecification}{bias}
Let $\bar m_a,\bar e$ be arbitrary (possibly wrong) limits of the nuisance estimators, with $\bar e$ bounded away from $0$ and $1$. Under \cref{ass:consistency,ass:exch,ass:pos,ass:sutva},
\[\E\{\psi(W;\bar\eta)\}-\ATE=\E\Big[\{\bar e(X)-e(X)\}\Big(\frac{\bar m_1(X)-m_1(X)}{\bar e(X)}+\frac{\bar m_0(X)-m_0(X)}{1-\bar e(X)}\Big)\Big].\]
Hence AIPW is unbiased if either $\bar e=e$ or $\bar m_a=m_a$ for both arms, and the bias is a \emph{product} of the two errors \citep{robins1994,bang2005,kang2007}.
\end{proposition}

\begin{proposition}{A robust nuisance fit cannot beat the efficiency bound}{eff}
Let $\sigma_a^2(x)=\Var(Y\mid A=a,X=x)<\infty$. The semiparametric efficiency bound for $\ATE$ in the model that imposes only \cref{ass:consistency,ass:exch,ass:pos,ass:sutva} is
\[V=\E\Big\{\frac{\sigma_1^2(X)}{e(X)}+\frac{\sigma_0^2(X)}{1-e(X)}+(\tau(X)-\ATE)^2\Big\},\qquad\tau(X)=m_1(X)-m_0(X)\]
\citep{hahn1998}. Cross-fitted AIPW whose nuisance estimators are consistent for the conditional \emph{mean} functions at product rate $o_p(n^{-1/2})$ is asymptotically linear with variance $V$ \citep{chernozhukov2018}, whichever consistent engine (OLS, Huber, $t$, $t$-mixture) produced $\hat m_a$.
\end{proposition}
\noindent\textit{Consequence.} Asymptotically, a robust outcome model cannot reduce the variance of the \emph{mean}-ATE estimator; contamination enters $V$ through $\sigma_a^2(x)$ whatever the engine. Any benefit must come from finite-sample behaviour, from misspecification or from a change of estimand. \Cref{sec:mc} quantifies each channel.

\begin{proposition}{Estimand implied by clipped augmentation}{clip}
Let $c(\cdot)$ be a bounded, odd, nondecreasing clipping function and suppose $\bar e=e$. Put $\delta_a(x)=\E\{c(Y-\bar m_a(x))\mid A=a,X=x\}$. Then
\[\E\{\psi_c(W;\bar\eta)\}=\E\big[\bar m_1(X)+\delta_1(X)-\bar m_0(X)-\delta_0(X)\big].\]
If the conditional law of $Y(a)-\bar m_a(X)$ is symmetric about $0$, then $\delta_a\equiv0$ and the clipped estimator is unbiased for $\ATE$ whenever in addition $\bar m_a=m_a$. More generally, if $\bar m_a(x)=\ell_a(x)$ is the $c$-location of $Y(a)\mid X=x$ (the root of $\E\{c(Y-\ell)\mid A=a,X=x\}=0$), then $\delta_a\equiv0$ and the estimator targets $\E\{\ell_1(X)-\ell_0(X)\}$, which differs from $\ATE$ unless the asymmetry of the outcome law is the same in both arms.
\end{proposition}

\begin{proposition}{Monotone ascent of the constrained ECM algorithm}{em}
Every iteration of the ECM algorithm of \cref{sec:methods}, with or without the lower bound $\sigma_k\ge\underline\sigma$, leaves the observed-data log-likelihood
$\ell(\beta,\sigma,\pi)=\sum_i\log\sum_k\pi_k\,t_\nu(y_i;\beta^\top D_i,\sigma_k^2)$ non-decreasing (up to the negligible ridge $10^{-8}$ used for numerical stability). Under standard regularity conditions its limit points are stationary points of $\ell$ on the constrained parameter set \citep{dempster1977,meng1993}.
\end{proposition}

\begin{remarkbox}
\textbf{Mean versus location.} A Student-$t$ regression estimates a conditional \emph{location}, which equals the conditional mean only if the error law is symmetric. Under asymmetric contamination $\hat m_a$ is therefore a misspecified mean model; by \cref{prop:bias} this is harmless when $\bar e=e$ (the augmentation corrects it) and costs variance, not consistency.
\end{remarkbox}
\begin{remarkbox}
\textbf{Moments.} A $t_\nu$ variable has a mean only for $\nu>1$ and a variance only for $\nu>2$. For $\nu\le2$, \cref{prop:eff} and the Wald interval \eqref{eq:se} do not apply; experiment E2 includes $\nu=1.5$ to show what happens when they fail.
\end{remarkbox}

\section{Monte Carlo design}\label{sec:design}
\subsection{Simulation model}
\begin{figure}[!htb]\centering\includegraphics[width=\textwidth]{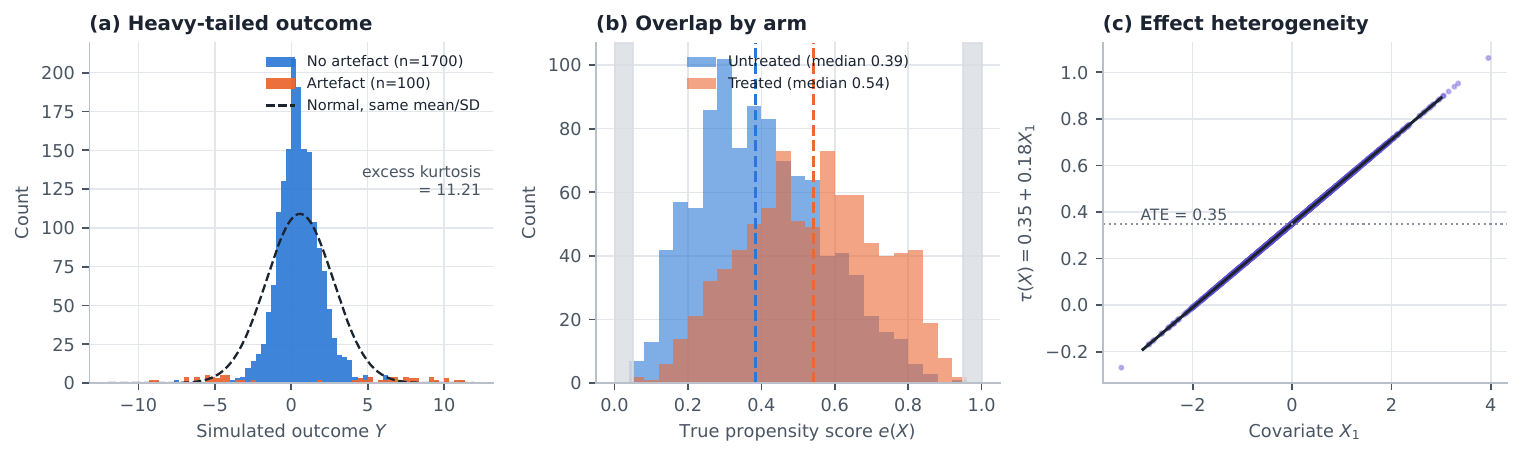}
\caption{One draw from the simulation model ($n=\nSim$, \contSim\% artefacts). (a) Outcome distribution, with artefact-affected observations in orange and a normal density with the same mean and SD for reference. (b) True propensity scores by arm; dashed lines are medians and grey bands the truncation region. (c) True conditional effect $\tau(X)=0.35+0.18X_1$, whose average is the target ATE $=0.35$.}\label{fig:simdesign}\end{figure}

Covariates $X=(X_1,X_2,X_3)^\top$ have independent $N(0,1)$ components. The propensity and potential outcomes are
\begin{align}
e(X)&=\expit(-0.2+0.7X_1-0.45X_2+0.3X_3),\qquad A\sim\mathrm{Bernoulli}\{e(X)\},\\
Y&=0.3+0.65X_1-0.3X_2+0.2X_3+A\,\tau(X)+\varepsilon+S\,\mathbb 1\{\text{artefact}\},\qquad \tau(X)=0.35+0.18X_1,
\end{align}
so that $\ATE=0.35$. The base error is a two-group Student-$t$ scale mixture: with probability $0.32$, $\varepsilon\sim1.45\,t_{3.5}$, otherwise $\varepsilon\sim0.55\,t_9$ (error variance $\approx1.84$). An \emph{artefact} adds a shift $S=\pm U(4,9)$ to a fraction of observations (symmetric) or $+U(4,9)$ (one-sided), either independently of $(X,A)$ or with arm-specific probabilities. It stands for a recording error and is \emph{not} part of the target outcome, so $\ATE=0.35$ in every scenario. \Cref{fig:simdesign} shows one draw of size $n=\nSim$.

\subsection{Experiments}
\begin{table}[!htb]\centering\small
\caption{Experiments. $B$ is the number of independent replications per cell; within a cell all seven estimators are applied to the same samples.}\label{tab:experiments}
\begin{tabular}{@{}llp{6.4cm}rr@{}}
\toprule\rowcolor{rowshade}\textbf{ID}&\textbf{Question}&\textbf{Design}&$\bm n$&$\bm B$\\\midrule
E1&Contamination&symmetric artefact probability $0,0.03,0.06,0.12,0.20$&650&\Breps\\
E2&Tail index&i.i.d.\ $t_\nu$ errors, $\nu\in\{1.5,2,3,5,10,30\}$, no artefacts&650&\BrepsOther\\
E3&Sample size&$n\in\{200,400,800,1600,3200\}$, artefact probability $0.06$&varies&\BrepsOther\\
E4&Misspecification&$2\times2$: outcome mean and/or propensity logit contain nonlinear terms unknown to the models&650&\BrepsOther\\
E5&Artefact structure&symmetric/one-sided $\times$ equal rate ($6\%$) / arm-specific ($3\%$ control, $9\%$ treated)&650&\BrepsOther\\
E6&Plasmode (DRC)&real covariates and drug packages of the \nApp{} DRC rounds; effect $\plasTau$ points; residuals resampled from the DRC fits, common or arm-specific&\nApp&\Bplas\\
E7&ECM diagnostics&likelihood traces and monotonicity over \emNfits{} simulated fits, plus the DRC fits&650&--\\
E8&Computing time&engines and full pipeline, $n$ from 500 to \timeNmax&varies&--\\\bottomrule
\end{tabular}\end{table}
Seeds are deterministic functions of the experiment, cell and replication index, so every number can be reproduced exactly.

\paragraph{Plasmode design (E6).} Following \citet{franklin2014}, E6 keeps everything about the DRC rounds that is not the outcome: the covariate matrix, the observed drug package and the zone structure. Outcomes are generated as $Y^\ast_i=\hat m_0(X_i)+\plasTau\,A_i+\varepsilon^\ast_i$, where $\hat m_0$ is the OLS fit of programme coverage among PZQ-only rounds and $\varepsilon^\ast_i$ is drawn with replacement from the centred OLS residuals of the real data, either pooled over both arms (\emph{common law}) or within the round's own arm (\emph{arm-specific laws}). Both residual laws have mean zero in each arm, so the true ATE is exactly $\plasTau$ points in both scenarios, but in the second the residuals are skewed differently in the two arms (skewness \plSkewZero{} for PZQ alone and \plSkewOne{} for integrated rounds). The estimators run the full application pipeline: five zone-level folds and cluster-robust standard errors.

\subsection{Performance measures}
For replication $b$ with estimate $\hat\tau_b$ and error $d_b=\hat\tau_b-\ATE$ we report \emph{bias} $\bar d$, \emph{empirical SD}, \emph{RMSE} $\{B^{-1}\sum_bd_b^2\}^{1/2}$, the mean estimated standard error, \emph{SE calibration} (mean $\widehat{\mathrm{SE}}$ divided by the empirical SD) and the \emph{coverage} of nominal 95\% Wald intervals. Every figure shows $\pm1.96$ Monte Carlo standard errors (MCSE), computed as in \citet{morris2019}: $\mathrm{sd}(d)/\sqrt B$ for bias and $\sqrt{\hat p(1-\hat p)/B}$ for coverage. The G-formula and IPW are reported without standard errors, because \eqref{eq:se} is the AIPW influence-function estimate and does not apply to them.

\section{Monte Carlo results}\label{sec:mc}
All numbers in this section are produced by the code from the stored simulation output.

\subsection{Contamination (E1)}
\begin{figure}[!htb]\centering\includegraphics[width=.84\textwidth]{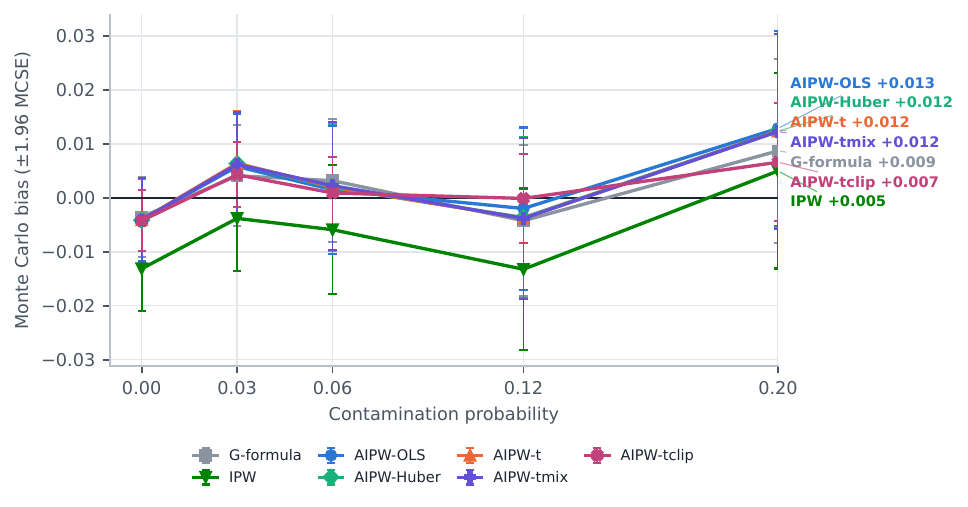}
\caption{Monte Carlo bias against contamination probability (E1, $n=650$, $B=\Breps$). Whiskers are $\pm1.96$ MCSE; labels give the value at 20\%. All seven estimators are unbiased within Monte Carlo error.}\label{fig:bias}\end{figure}

\begin{figure}[!htb]\centering\includegraphics[width=.84\textwidth]{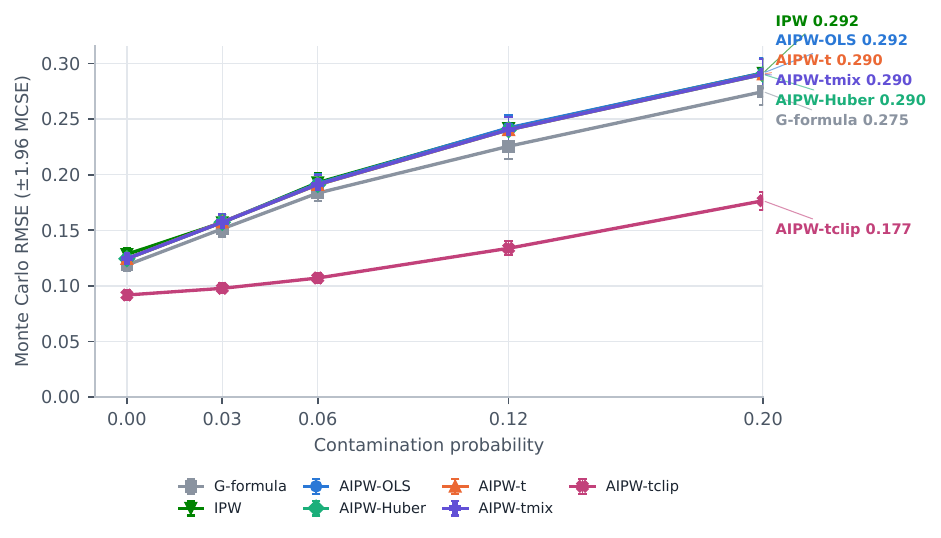}
\caption{RMSE against contamination probability (E1). The four unclipped AIPW variants coincide; AIPW-$t$clip is far lower. The G-formula does slightly better than AIPW because its outcome model is correctly specified here.}\label{fig:rmse}\end{figure}

\begin{figure}[!htb]\centering\includegraphics[width=.84\textwidth]{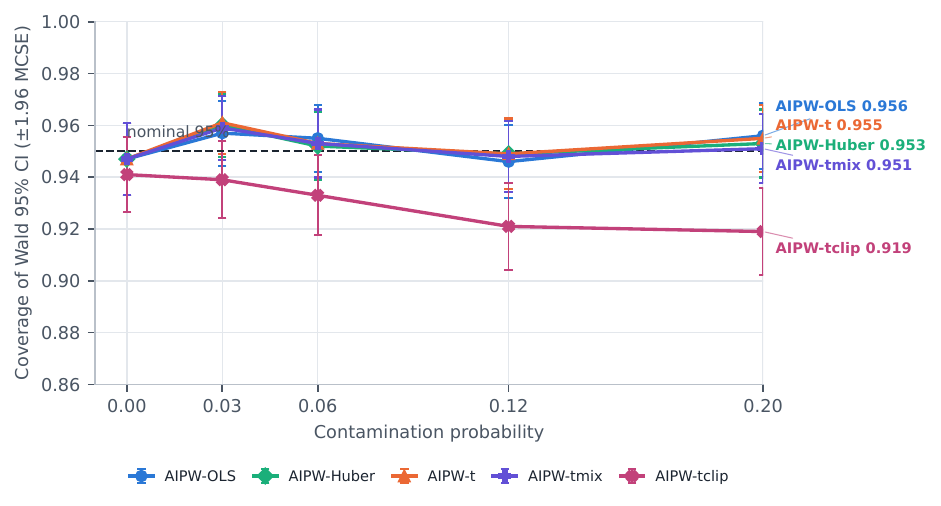}
\caption{Coverage of nominal 95\% Wald intervals (E1; dashed line: nominal level). Unclipped AIPW variants are well calibrated; clipped AIPW under-covers because its standard error is too small.}\label{fig:cov}\end{figure}

\begin{figure}[!htb]\centering\includegraphics[width=.84\textwidth]{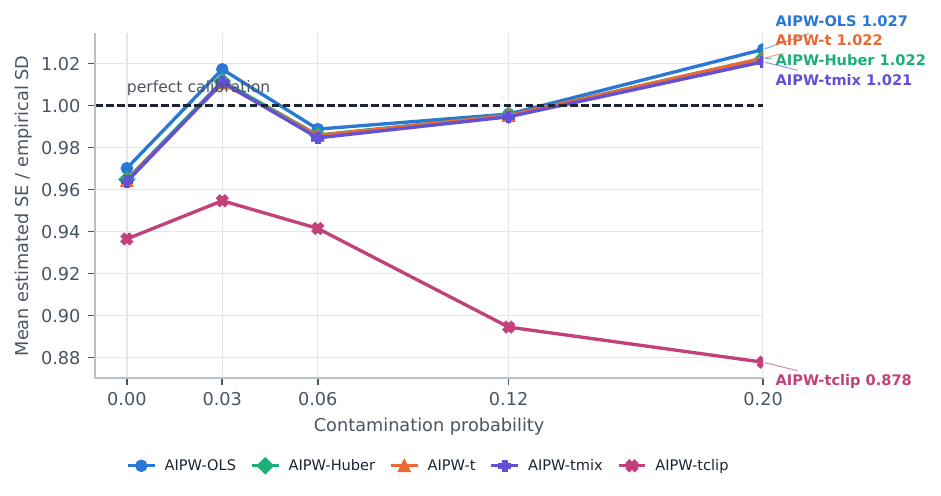}
\caption{Calibration of the standard error (E1): mean estimated SE divided by the empirical SD of the estimates; 1 means perfect calibration.}\label{fig:se}\end{figure}

\begin{figure}[!htb]\centering\includegraphics[width=.84\textwidth]{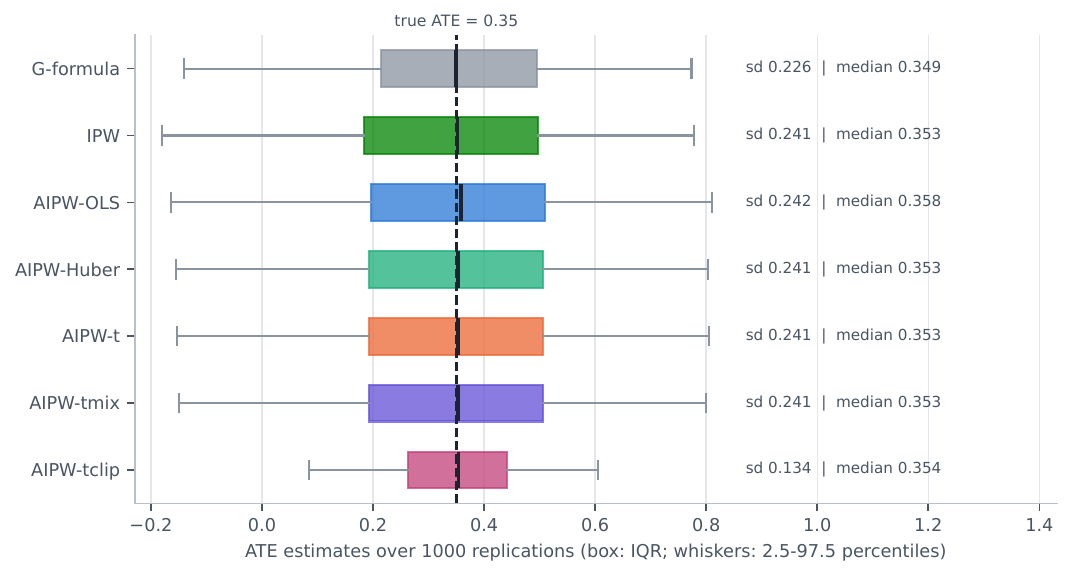}
\caption{Sampling distributions of the seven estimators at 12\% contamination (E1, $B=\Breps$). Boxes: interquartile range; whiskers: 2.5--97.5 percentiles; text: SD and median. The dashed line is the true ATE.}\label{fig:dist}\end{figure}

\begin{figure}[!htb]\centering\includegraphics[width=.76\textwidth]{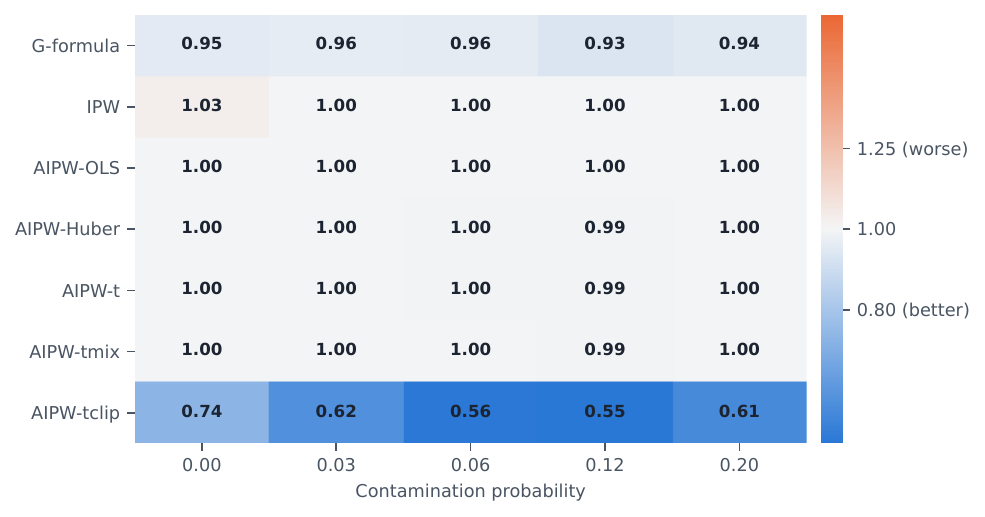}
\caption{RMSE relative to AIPW-OLS (E1). Blue cells beat AIPW-OLS, orange cells do worse; ratios are printed in the cells.}\label{fig:relrmse}\end{figure}

\Cref{tab:contamination} and \cref{fig:bias,fig:rmse,fig:cov,fig:se,fig:dist,fig:relrmse} summarize the main experiment. Four findings stand out.

\paragraph{No estimator is meaningfully biased.} All biases are within \maxAbsBias{} in absolute value, with MCSE between \mcseLo{} and \mcseHi.

\paragraph{Robust outcome engines change nothing.} AIPW-Huber, AIPW-$t$ and AIPW-$t$mix are indistinguishable from AIPW-OLS: at 6\% contamination their RMSEs are \rmseHuberSix, \rmseTSix{} and \rmseTmixSix, against \rmseOlsSix{} for OLS, and the largest RMSE difference to OLS in any cell is \maxRmseDiff{} (\cref{fig:rmse,fig:relrmse}). This is what \cref{prop:eff} predicts. The empirical SDs also match the asymptotic values $\sqrt{V/n}$ computed from the simulation model (\asdZero, \asdSix, \asdTwelve, \asdTwenty{} at contamination $0$, $0.06$, $0.12$, $0.20$, against \sdOlsZero, \sdOlsSix, \sdOlsTwelve, \sdOlsTwenty{} observed for AIPW-OLS); the small excess plausibly reflects finite $n$ and propensity truncation.

\paragraph{Clipping the augmentation helps a lot.} AIPW-$t$clip has the lowest RMSE of the doubly robust estimators in every cell: \rmseClipZero{} versus \rmseOlsZero{} without artefacts, \rmseClipSix{} versus \rmseOlsSix{} at 6\%, and \rmseClipTwenty{} versus \rmseOlsTwenty{} at 20\%, a reduction of \clipRedLo--\clipRedHi\%. Because the artefacts here are symmetric, \cref{prop:clip} guarantees that the gain costs no bias. Part of it, even without artefacts, comes from the heavy-tailed base errors.

\paragraph{Calibration.} Wald intervals for the unclipped AIPW variants cover between \covUncLo{} and \covUncHi{} in all cells, with SE ratios near one. For clipped AIPW the standard error ignores the estimation of the clipping scale and is too small: coverage falls from \covClipZero{} to \covClipTwenty{} and the SE ratio to \seratioClipTwenty{} at 20\% contamination. A bootstrap, or a variance estimator that accounts for the clipping step, is needed before clipped intervals are reported.


\needspace{14\baselineskip}
{\small\setlength{\LTcapwidth}{\textwidth}
\begin{longtable}{@{}lrrrrrr@{}}
\caption{Contamination sweep (E1, $n=650$, $B=\Breps$). MCSE is the Monte Carlo standard error of the bias; the best RMSE in each block is in bold. Standard errors and coverage are not defined for the G-formula and IPW (``--).}\label{tab:contamination}\\
\toprule\rowcolor{rowshade}\textbf{Method}&\textbf{Bias}&\textbf{MCSE}&\textbf{SD}&\textbf{RMSE}&\textbf{Mean SE}&\textbf{Coverage}\\\midrule\endfirsthead
\toprule\rowcolor{rowshade}\textbf{Method}&\textbf{Bias}&\textbf{MCSE}&\textbf{SD}&\textbf{RMSE}&\textbf{Mean SE}&\textbf{Coverage}\\\midrule\endhead
\rowcolor{rowshade}\textbf{Contamination probability 0.00}&&&&&&\\
G-formula & -0.004 & 0.004 & 0.119 & 0.119 & -- & --\\
IPW & -0.013 & 0.004 & 0.128 & 0.128 & -- & --\\
AIPW-OLS & -0.004 & 0.004 & 0.124 & 0.124 & 0.121 & 0.947\\
AIPW-Huber & -0.004 & 0.004 & 0.124 & 0.124 & 0.120 & 0.947\\
AIPW-$t$ & -0.004 & 0.004 & 0.124 & 0.124 & 0.120 & 0.947\\
AIPW-$t$mix & -0.004 & 0.004 & 0.124 & 0.124 & 0.120 & 0.947\\
AIPW-$t$clip & -0.004 & 0.003 & 0.092 & \textbf{0.092} & 0.086 & 0.941\\
\rowcolor{rowshade}\textbf{Contamination probability 0.03}&&&&&&\\
G-formula & +0.004 & 0.005 & 0.151 & 0.151 & -- & --\\
IPW & -0.004 & 0.005 & 0.157 & 0.157 & -- & --\\
AIPW-OLS & +0.006 & 0.005 & 0.157 & 0.157 & 0.160 & 0.957\\
AIPW-Huber & +0.006 & 0.005 & 0.157 & 0.157 & 0.159 & 0.960\\
AIPW-$t$ & +0.006 & 0.005 & 0.157 & 0.157 & 0.159 & 0.961\\
AIPW-$t$mix & +0.006 & 0.005 & 0.157 & 0.157 & 0.159 & 0.959\\
AIPW-$t$clip & +0.004 & 0.003 & 0.098 & \textbf{0.098} & 0.093 & 0.939\\
\rowcolor{rowshade}\textbf{Contamination probability 0.06}&&&&&&\\
G-formula & +0.003 & 0.006 & 0.184 & 0.183 & -- & --\\
IPW & -0.006 & 0.006 & 0.193 & 0.193 & -- & --\\
AIPW-OLS & +0.001 & 0.006 & 0.192 & 0.192 & 0.190 & 0.955\\
AIPW-Huber & +0.002 & 0.006 & 0.191 & 0.191 & 0.188 & 0.952\\
AIPW-$t$ & +0.002 & 0.006 & 0.191 & 0.191 & 0.188 & 0.953\\
AIPW-$t$mix & +0.002 & 0.006 & 0.191 & 0.191 & 0.188 & 0.953\\
AIPW-$t$clip & +0.001 & 0.003 & 0.107 & \textbf{0.107} & 0.101 & 0.933\\
\rowcolor{rowshade}\textbf{Contamination probability 0.12}&&&&&&\\
G-formula & -0.004 & 0.007 & 0.226 & 0.226 & -- & --\\
IPW & -0.013 & 0.008 & 0.241 & 0.242 & -- & --\\
AIPW-OLS & -0.002 & 0.008 & 0.242 & 0.242 & 0.241 & 0.946\\
AIPW-Huber & -0.004 & 0.008 & 0.241 & 0.240 & 0.239 & 0.949\\
AIPW-$t$ & -0.004 & 0.008 & 0.241 & 0.240 & 0.239 & 0.949\\
AIPW-$t$mix & -0.004 & 0.008 & 0.241 & 0.241 & 0.239 & 0.948\\
AIPW-$t$clip & -0.000 & 0.004 & 0.134 & \textbf{0.134} & 0.120 & 0.921\\
\rowcolor{rowshade}\textbf{Contamination probability 0.20}&&&&&&\\
G-formula & +0.009 & 0.009 & 0.275 & 0.275 & -- & --\\
IPW & +0.005 & 0.009 & 0.292 & 0.292 & -- & --\\
AIPW-OLS & +0.013 & 0.009 & 0.291 & 0.292 & 0.299 & 0.956\\
AIPW-Huber & +0.012 & 0.009 & 0.290 & 0.290 & 0.297 & 0.953\\
AIPW-$t$ & +0.012 & 0.009 & 0.290 & 0.290 & 0.297 & 0.955\\
AIPW-$t$mix & +0.012 & 0.009 & 0.290 & 0.290 & 0.296 & 0.951\\
AIPW-$t$clip & +0.007 & 0.006 & 0.177 & \textbf{0.177} & 0.155 & 0.919\\
\bottomrule
\end{longtable}}

\subsection{Tail index (E2)}
\begin{figure}[!htb]\centering\includegraphics[width=\textwidth]{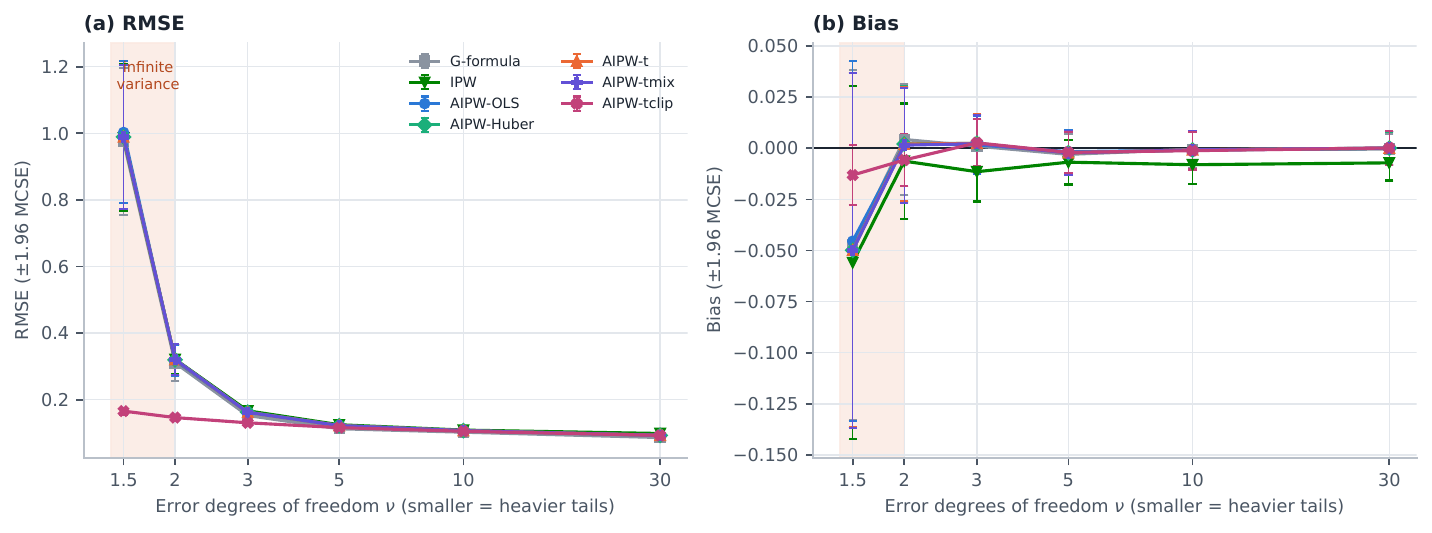}
\caption{Effect of the tail index (E2, $n=650$, no artefacts, $B=\BrepsOther$). (a) RMSE; (b) bias with $\pm1.96$ MCSE whiskers. The shaded band marks $\nu\le2$, where the error variance is infinite.}\label{fig:tails}\end{figure}

\Cref{fig:tails} and \cref{tab:tails} replace the contaminated mixture by i.i.d.\ $t_\nu$ errors. For $\nu\ge3$ the unclipped AIPW variants coincide (for instance \tailThreeOls{} for OLS and \tailThreeT{} for $t$ at $\nu=3$), and the advantage of clipping fades for light tails ($\nu=30$: \tailThirtyDfOls{} against \tailThirtyDfClip). The picture changes at $\nu\le2$, where the variance is infinite and \cref{prop:eff} no longer applies. At $\nu=1.5$ the RMSE of the unclipped estimators jumps to \tailOneFiveOls{} for OLS and \tailOneFiveT{} for $t$; the Huber and $t$ engines bound $\hat m_a$ but not the augmentation, so they do not help, whereas the clipped estimator keeps an RMSE of \tailOneFiveClip, about \tailFactor{} times smaller. Coverage of the unclipped intervals means little there (SE ratio \tailSeRatioOneFive{} yet coverage \tailCovOneFive), one more reason for caution with infinite-variance outcomes.
\begin{table}[!htb]\centering\small
\caption{RMSE by error degrees of freedom (E2, $n=650$, $B=\BrepsOther$) and coverage of the OLS and $t$ AIPW intervals. Best RMSE in bold.}\label{tab:tails}
\begin{tabular}{@{}rrrrrrrr@{}}\toprule
\rowcolor{rowshade}$\bm\nu$&\textbf{AIPW-OLS}&\textbf{AIPW-Huber}&\textbf{AIPW-$t$}&\textbf{AIPW-$t$mix}&\textbf{AIPW-$t$clip}&\textbf{Cov.\ OLS}&\textbf{Cov.\ $t$}\\\midrule
1.5 & 1.003 & 0.989 & 0.989 & 0.989 & \textbf{0.166} & 0.974 & 0.970\\
2 & 0.318 & 0.320 & 0.320 & 0.320 & \textbf{0.146} & 0.966 & 0.956\\
3 & 0.163 & 0.162 & 0.162 & 0.162 & \textbf{0.131} & 0.948 & 0.944\\
5 & 0.122 & 0.122 & 0.122 & 0.122 & \textbf{0.116} & 0.932 & 0.932\\
10 & 0.107 & 0.107 & 0.107 & 0.107 & \textbf{0.105} & 0.942 & 0.942\\
30 & \textbf{0.093} & 0.093 & 0.093 & 0.093 & 0.093 & 0.956 & 0.962\\
\bottomrule\end{tabular}\end{table}

\subsection{Sample size (E3)}
\begin{figure}[!htb]\centering\includegraphics[width=.84\textwidth]{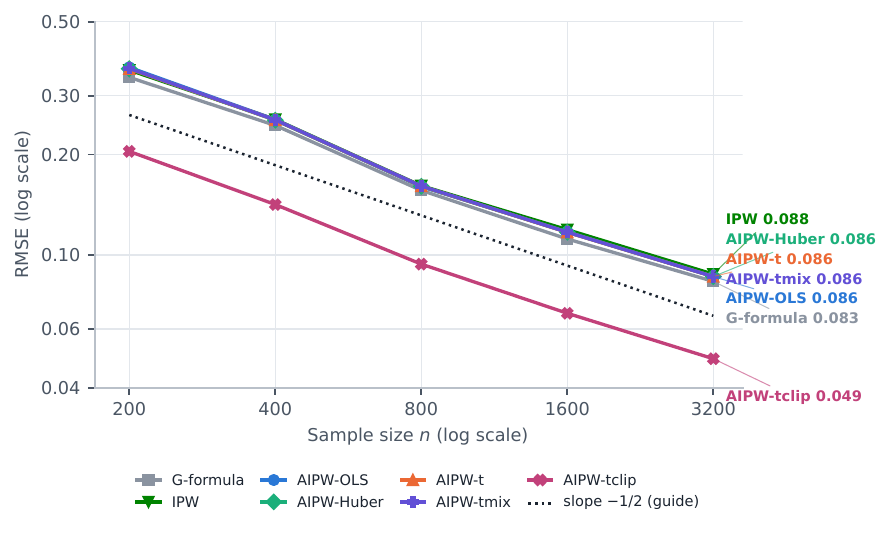}
\caption{RMSE against sample size on log--log axes (E3, 6\% artefacts, $B=\BrepsOther$). Dotted line: slope $-1/2$; labels give the RMSE at $n=3200$.}\label{fig:size}\end{figure}

RMSE falls at the $n^{-1/2}$ rate for both the unclipped and the clipped estimator (fitted log--log slopes \slopeOls{} and \slopeClip; \cref{fig:size,tab:size}), and the ratio of clipped to unclipped RMSE stays between \sizeRatioLo{} and \sizeRatioHi. Coverage of the clipped interval does not approach 0.95 as $n$ grows (it ranges from \sizeCovClipLo{} to \sizeCovClipHi), which confirms that the shortfall belongs to the variance estimator and not to small samples.
\begin{table}[!htb]\centering\small
\caption{RMSE and coverage by sample size (E3, 6\% artefacts, $B=\BrepsOther$).}\label{tab:size}
\begin{tabular}{@{}rrrrrrrr@{}}\toprule
\rowcolor{rowshade}$\bm n$&\textbf{RMSE OLS}&\textbf{RMSE $t$}&\textbf{RMSE $t$mix}&\textbf{RMSE $t$clip}&\textbf{Cov.\ OLS}&\textbf{Cov.\ $t$}&\textbf{Cov.\ $t$clip}\\\midrule
200 & 0.365 & 0.361 & 0.362 & 0.204 & 0.964 & 0.952 & 0.946\\
400 & 0.254 & 0.254 & 0.254 & 0.142 & 0.944 & 0.952 & 0.924\\
800 & 0.160 & 0.161 & 0.161 & 0.094 & 0.962 & 0.962 & 0.938\\
1600 & 0.117 & 0.117 & 0.117 & 0.067 & 0.956 & 0.958 & 0.930\\
3200 & 0.086 & 0.086 & 0.086 & 0.049 & 0.942 & 0.944 & 0.924\\
\bottomrule\end{tabular}\end{table}

\subsection{Nuisance misspecification (E4)}
\begin{figure}[!htb]\centering\includegraphics[width=.9\textwidth]{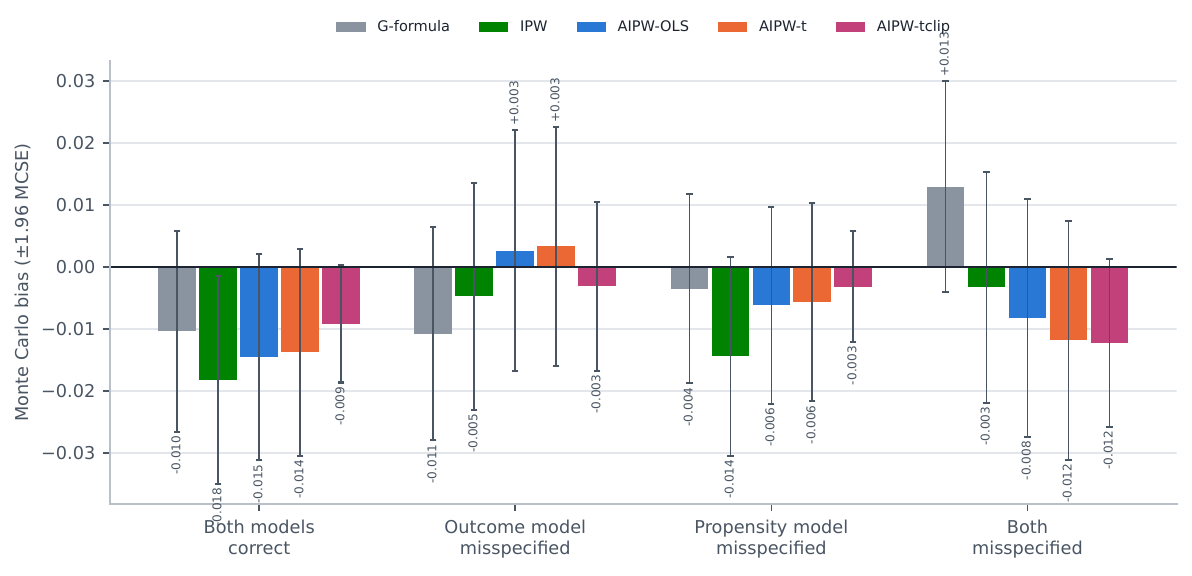}
\caption{Bias under outcome and/or propensity misspecification (E4, $n=650$, 6\% artefacts, $B=\BrepsOther$). Whiskers: $\pm1.96$ MCSE; rotated labels: values.}\label{fig:misspec}\end{figure}

With the nonlinear terms of E4 left out of the working models, every estimator stays essentially unbiased (largest absolute bias \misMaxAbs; \cref{fig:misspec,tab:misspec}); only the variance of unclipped AIPW grows (RMSE \misRmseOutBad{} when the outcome model is wrong, against \misRmseOk{} when both are right). \emph{This experiment is too mild to show double robustness empirically}: the biases of the G-formula and of IPW are also within Monte Carlo error, so the data cannot separate the estimators on bias. Stronger misspecification would be needed for that, and its absence is a limitation of the study; \cref{prop:bias} gives the exact bias formula.
\begin{table}[!htb]\centering\small
\caption{Misspecification experiment (E4, $n=650$, 6\% artefacts, $B=\BrepsOther$).}\label{tab:misspec}
\begin{tabular}{@{}lrrrr@{}}\toprule
\rowcolor{rowshade}\textbf{Method}&\textbf{Bias}&\textbf{MCSE}&\textbf{RMSE}&\textbf{Coverage}\\\midrule
\rowcolor{rowshade}\textbf{Both correct}&&&&\\
G-formula & -0.010 & 0.008 & 0.185 & --\\
IPW & -0.018 & 0.009 & 0.192 & --\\
AIPW-OLS & -0.015 & 0.008 & 0.190 & 0.952\\
AIPW-$t$ & -0.014 & 0.009 & 0.191 & 0.944\\
AIPW-$t$clip & -0.009 & 0.005 & 0.108 & 0.930\\
\rowcolor{rowshade}\textbf{Outcome model misspecified}&&&&\\
G-formula & -0.011 & 0.009 & 0.197 & --\\
IPW & -0.005 & 0.009 & 0.208 & --\\
AIPW-OLS & +0.003 & 0.010 & 0.221 & 0.960\\
AIPW-$t$ & +0.003 & 0.010 & 0.220 & 0.958\\
AIPW-$t$clip & -0.003 & 0.007 & 0.155 & 0.942\\
\rowcolor{rowshade}\textbf{Propensity model misspecified}&&&&\\
G-formula & -0.004 & 0.008 & 0.174 & --\\
IPW & -0.014 & 0.008 & 0.183 & --\\
AIPW-OLS & -0.006 & 0.008 & 0.181 & 0.958\\
AIPW-$t$ & -0.006 & 0.008 & 0.182 & 0.958\\
AIPW-$t$clip & -0.003 & 0.005 & 0.102 & 0.938\\
\rowcolor{rowshade}\textbf{Both misspecified}&&&&\\
G-formula & +0.013 & 0.009 & 0.195 & --\\
IPW & -0.003 & 0.009 & 0.212 & --\\
AIPW-OLS & -0.008 & 0.010 & 0.218 & 0.942\\
AIPW-$t$ & -0.012 & 0.010 & 0.220 & 0.942\\
AIPW-$t$clip & -0.012 & 0.007 & 0.155 & 0.928\\
\bottomrule\end{tabular}\end{table}

\subsection{Artefact structure (E5): where no method is safe}
\begin{figure}[!htb]\centering\includegraphics[width=\textwidth]{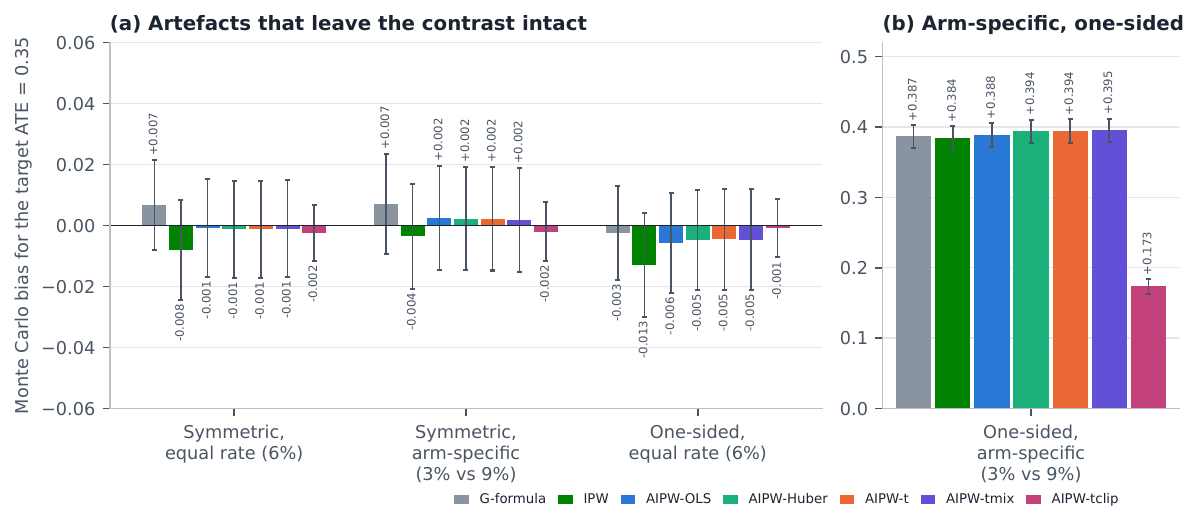}
\caption{Bias for the target ATE $=0.35$ under four artefact structures (E5, $n=650$, $B=\BrepsOther$). (a) Three structures that leave the contrast intact (note the narrow axis, $\pm0.06$). (b) One-sided artefacts at $3\%$ (control) and $9\%$ (treated). Whiskers: $\pm1.96$ MCSE; rotated labels: values.}\label{fig:artefacts}\end{figure}

\Cref{fig:artefacts,tab:artefacts} contrast four artefact structures with the same average rate. Symmetric artefacts, whether equal or arm-specific, and one-sided artefacts at equal rates leave every estimator unbiased (largest absolute bias \artOkMax). \emph{One-sided artefacts at different rates in the two arms ($3\%$ against $9\%$) bias every estimator}, by roughly $(0.09-0.03)\times6.5\approx0.39$, the difference in mean artefact shifts: G-formula \artPosDiffGcomp, IPW \artPosDiffIpw, AIPW-OLS \artPosDiffOls, AIPW-Huber \artPosDiffHuber, AIPW-$t$ \artPosDiffT, AIPW-$t$mix \artPosDiffTmix. Robust outcome models do not reduce the bias, because the augmentation term brings the raw residuals back (\cref{prop:bias,prop:eff}). Clipping reduces it to \artPosDiffClip{} without removing it, since clipping at three robust standard deviations truncates the artefacts rather than discarding them. Coverage is then poor for every method (\artCovPosDiffOls{} for AIPW-OLS, \artCovPosDiffClip{} for clipped AIPW). Differential recording artefacts cannot be repaired by choosing a robust regression; they have to be traced to the way the data are produced and handled by design or by sensitivity analysis. \Cref{sec:app} meets exactly this situation in the DRC data.
\needspace{14\baselineskip}
{\small\setlength{\LTcapwidth}{\textwidth}
\begin{longtable}{@{}lrrrrr@{}}
\caption{Artefact structures (E5, $n=650$, $B=\BrepsOther$); target ATE $=0.35$.}\label{tab:artefacts}\\
\toprule\rowcolor{rowshade}\textbf{Method}&\textbf{Bias}&\textbf{MCSE}&\textbf{SD}&\textbf{RMSE}&\textbf{Coverage}\\\midrule\endfirsthead
\toprule\rowcolor{rowshade}\textbf{Method}&\textbf{Bias}&\textbf{MCSE}&\textbf{SD}&\textbf{RMSE}&\textbf{Coverage}\\\midrule\endhead
\rowcolor{rowshade}\textbf{Symmetric, equal rate}&&&&&\\
G-formula & +0.007 & 0.008 & 0.168 & 0.168 & --\\
IPW & -0.008 & 0.008 & 0.187 & 0.187 & --\\
AIPW-OLS & -0.001 & 0.008 & 0.183 & 0.183 & 0.960\\
AIPW-Huber & -0.001 & 0.008 & 0.182 & 0.182 & 0.966\\
AIPW-$t$ & -0.001 & 0.008 & 0.182 & 0.182 & 0.966\\
AIPW-$t$mix & -0.001 & 0.008 & 0.181 & 0.181 & 0.966\\
AIPW-$t$clip & -0.002 & 0.005 & 0.105 & 0.105 & 0.938\\
\rowcolor{rowshade}\textbf{Symmetric, differing by arm}&&&&&\\
G-formula & +0.007 & 0.008 & 0.187 & 0.187 & --\\
IPW & -0.004 & 0.009 & 0.197 & 0.196 & --\\
AIPW-OLS & +0.002 & 0.009 & 0.195 & 0.195 & 0.962\\
AIPW-Huber & +0.002 & 0.009 & 0.194 & 0.194 & 0.966\\
AIPW-$t$ & +0.002 & 0.009 & 0.194 & 0.194 & 0.966\\
AIPW-$t$mix & +0.002 & 0.009 & 0.194 & 0.194 & 0.962\\
AIPW-$t$clip & -0.002 & 0.005 & 0.110 & 0.109 & 0.930\\
\rowcolor{rowshade}\textbf{One-sided, equal rate}&&&&&\\
G-formula & -0.003 & 0.008 & 0.175 & 0.175 & --\\
IPW & -0.013 & 0.009 & 0.194 & 0.195 & --\\
AIPW-OLS & -0.006 & 0.008 & 0.188 & 0.188 & 0.954\\
AIPW-Huber & -0.005 & 0.008 & 0.188 & 0.187 & 0.958\\
AIPW-$t$ & -0.005 & 0.008 & 0.188 & 0.188 & 0.960\\
AIPW-$t$mix & -0.005 & 0.008 & 0.188 & 0.187 & 0.960\\
AIPW-$t$clip & -0.001 & 0.005 & 0.109 & 0.109 & 0.934\\
\rowcolor{rowshade}\textbf{One-sided, differing by arm}&&&&&\\
G-formula & +0.387 & 0.008 & 0.182 & 0.427 & --\\
IPW & +0.384 & 0.009 & 0.194 & 0.430 & --\\
AIPW-OLS & +0.388 & 0.009 & 0.194 & 0.434 & 0.468\\
AIPW-Huber & +0.394 & 0.009 & 0.192 & 0.438 & 0.454\\
AIPW-$t$ & +0.394 & 0.009 & 0.192 & 0.438 & 0.460\\
AIPW-$t$mix & +0.395 & 0.009 & 0.192 & 0.439 & 0.452\\
AIPW-$t$clip & +0.173 & 0.005 & 0.117 & 0.209 & 0.614\\
\bottomrule
\end{longtable}}

\subsection{Plasmode experiment on the DRC design (E6)}
\begin{figure}[!htb]\centering\includegraphics[width=\textwidth]{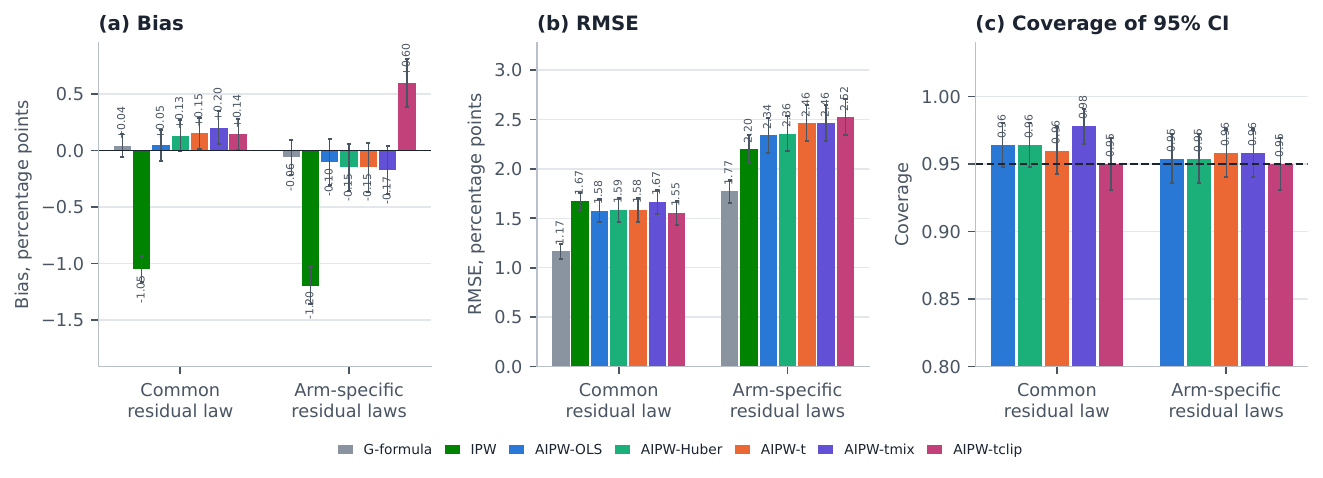}
\caption{Plasmode experiment E6 on the \nApp{} DRC rounds (true effect $\plasTau$ percentage points, $B=\Bplas$). (a) Bias, (b) RMSE, (c) coverage of 95\% Wald intervals with cluster-robust standard errors (AIPW only; dashed line: nominal level). Left groups: residuals drawn from a common law; right groups: from arm-specific laws. Whiskers: $\pm1.96$ MCSE.}\label{fig:plasmode}\end{figure}

With a common residual law the doubly robust estimators recover the effect of \plasTau{} points to within a fraction of a point: the biases are \plPoolBiasOls{} (AIPW-OLS), \plPoolBiasT{} (AIPW-$t$) and \plPoolBiasClip{} (AIPW-$t$clip), and the cluster-robust intervals of the unclipped variants cover \plPoolCovOls{} (OLS) and \plPoolCovT{} ($t$) of the time (\cref{fig:plasmode,tab:plasmode}). IPW, in contrast, is off by \plPoolBiasIpw{} points: the logistic propensity model is evidently not exact for the real assignment mechanism, and only the estimators with a correct outcome model escape the consequences, as \cref{prop:bias} says. The small positive biases of the $t$ engines are of the same nature: on skewed residuals a $t$ fit estimates a location rather than a mean, and the imperfect propensity lets a little of that through. Clipping barely helps here (RMSE \plPoolRmseClip{} against \plPoolRmseOls{} for AIPW-OLS), because the real residuals are heavy in the centre rather than in the extremes. With arm-specific residual laws the unclipped estimators remain unbiased (\plArmBiasOls{} for OLS, \plArmBiasT{} for $t$), whereas the clipped estimator now targets a clipped-location contrast and is off by \plArmBiasClip{} points, about five Monte Carlo standard errors; its RMSE (\plArmRmseClip) is then worse than that of AIPW-OLS (\plArmRmseOls). The bias is small next to the standard error, so coverage is not visibly affected (\plArmCovClip), but it does not shrink with $n$. This is \cref{prop:clip} at work on residual shapes taken from real data, which are skewed in one arm (skewness \plSkewZero) and nearly symmetric in the other (\plSkewOne).
\needspace{14\baselineskip}
{\small\setlength{\LTcapwidth}{\textwidth}
\begin{longtable}{@{}lrrrrr@{}}
\caption{Plasmode study (E6): the $n=\nApp$ DRC rounds with their real covariates and drug packages, outcomes generated with a known effect of $\plasTau$ percentage points and residuals resampled from the fitted DRC residuals ($B=\Bplas$). Coverage uses cluster-robust standard errors.}\label{tab:plasmode}\\
\toprule\rowcolor{rowshade}\textbf{Method}&\textbf{Bias}&\textbf{MCSE}&\textbf{SD}&\textbf{RMSE}&\textbf{Coverage}\\\midrule\endfirsthead
\toprule\rowcolor{rowshade}\textbf{Method}&\textbf{Bias}&\textbf{MCSE}&\textbf{SD}&\textbf{RMSE}&\textbf{Coverage}\\\midrule\endhead
\rowcolor{rowshade}\textbf{Common residual law}&&&&&\\
G-formula & +0.04 & 0.05 & 1.17 & 1.17 & --\\
IPW & -1.05 & 0.06 & 1.30 & 1.67 & --\\
AIPW-OLS & +0.05 & 0.07 & 1.58 & 1.58 & 0.964\\
AIPW-Huber & +0.13 & 0.07 & 1.58 & 1.59 & 0.964\\
AIPW-$t$ & +0.15 & 0.07 & 1.58 & 1.58 & 0.960\\
AIPW-$t$mix & +0.20 & 0.07 & 1.66 & 1.67 & 0.978\\
AIPW-$t$clip & +0.14 & 0.07 & 1.55 & 1.55 & 0.950\\
\rowcolor{rowshade}\textbf{Arm-specific residual laws}&&&&&\\
G-formula & -0.06 & 0.08 & 1.77 & 1.77 & --\\
IPW & -1.20 & 0.08 & 1.85 & 2.20 & --\\
AIPW-OLS & -0.10 & 0.10 & 2.34 & 2.34 & 0.954\\
AIPW-Huber & -0.15 & 0.11 & 2.35 & 2.36 & 0.954\\
AIPW-$t$ & -0.15 & 0.11 & 2.46 & 2.46 & 0.958\\
AIPW-$t$mix & -0.17 & 0.11 & 2.46 & 2.46 & 0.958\\
AIPW-$t$clip & +0.60 & 0.11 & 2.45 & 2.52 & 0.950\\
\bottomrule
\end{longtable}}

\subsection{ECM diagnostics (E7)}\label{sec:mc_em}
\begin{figure}[!htb]\centering\includegraphics[width=\textwidth]{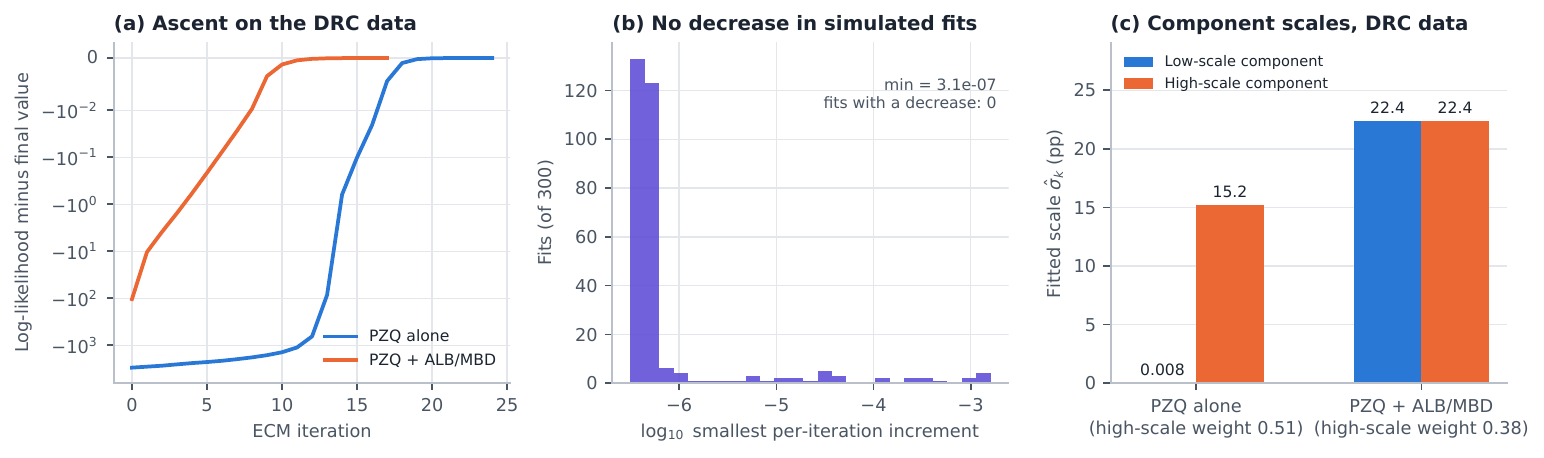}
\caption{ECM diagnostics. (a) Log-likelihood minus its final value for the arm-specific fits on the DRC programme-coverage data (symmetric-log axis). (b) Smallest per-iteration log-likelihood increment over \emNfits{} simulated fits at three contamination levels; no fit decreased. (c) Fitted component scales on the DRC data (percentage points), with the weight of the high-scale component under each arm.}\label{fig:em}\end{figure}

Over \emNfits{} simulated fits the observed-data log-likelihood never decreased (smallest per-iteration increment $\emMinInc$; \emNdec{} fits with a decrease beyond $10^{-8}$; median \emIterMed{} iterations), as \cref{prop:em} requires (\cref{fig:em}b). On the DRC programme-coverage data the algorithm also ascends monotonically in both arms (smallest increment $\emAppMinInc$; \emIterA{} and \emIterB{} iterations; \cref{fig:em}a), but the fitted components tell a different story (\cref{fig:em}c). Among PZQ-only rounds one component collapses onto the heaps and stops at the lower bound ($\hat\sigma=\emSigLowA$ points, against $\emSigHighA$ for the other component, weight \emPiA{}); among integrated rounds the two components end with almost the same scale ($\emSigLowB$ and $\emSigHighB$), that is, the mixture reduces to a single $t$ law. Neither fit describes a meaningful latent subpopulation. The mixture is a usable location engine on such data only because of the bound of \cref{sec:numerics}, and its components should not be interpreted.

\subsection{Computing time (E8)}\label{sec:timing}
\begin{figure}[!htb]\centering\includegraphics[width=\textwidth]{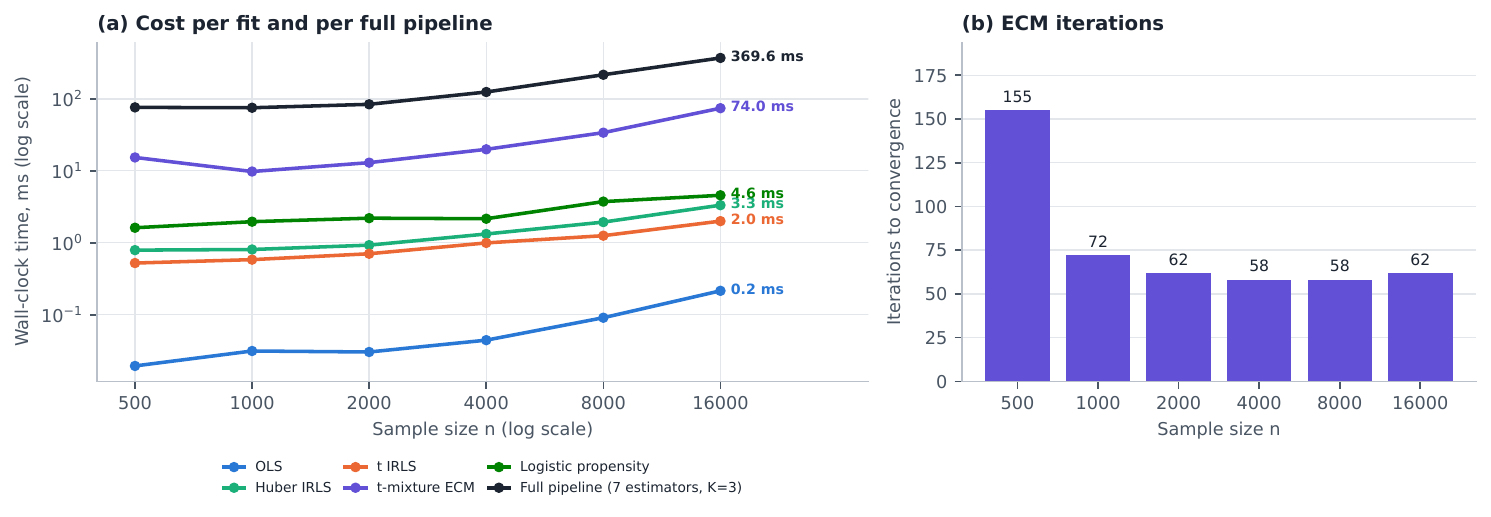}
\caption{Computing time (E8). (a) Median wall-clock time of one fit of each engine and of the full seven-estimator pipeline against $n$ (log--log; labels give the time at the largest $n$). (b) ECM iterations to convergence (relative tolerance $10^{-9}$).}\label{fig:timing}\end{figure}

\Cref{fig:timing,tab:timing} report median wall-clock times on one core (\machine). At $n=\timeNmax$ an OLS fit takes \timeOlsMax{} ms and an ECM fit \timeEcmMax{} ms, about \timeEcmRatio{} times longer; the complete pipeline with all seven estimators and three folds takes \timePipeSmall{} ms at $n=500$ and \timePipeMax{} s at $n=\timeNmax$. Fixed overheads dominate below a few thousand observations, and even from $n=2000$ upwards the growth is sub-linear over the range studied (log--log slope \timeSlope). The cost of a $t$ or mixture engine is therefore not a practical obstacle; a cluster bootstrap with a few hundred resamples of the DRC data runs in minutes.
\begin{table}[!htb]\centering\footnotesize
\caption{Median wall-clock time in milliseconds (E8, one core, x86\_64 / Python 3.13.18 / NumPy 2.5.3). Engines are fitted to the treated arm of a simulated sample of size $n$ (6\% artefacts); the full pipeline fits all nuisance models on three folds and returns the seven estimates.}\label{tab:timing}
\begin{tabular}{@{}lrrrrrr@{}}\toprule
\rowcolor{rowshade}\textbf{Task}&$\bm{n=500}$&$\bm{n=1000}$&$\bm{n=2000}$&$\bm{n=4000}$&$\bm{n=8000}$&$\bm{n=16000}$\\\midrule
OLS & 0.02 & 0.03 & 0.03 & 0.04 & 0.09 & 0.21\\
Huber IRLS & 0.79 & 0.81 & 0.93 & 1.3 & 1.9 & 3.3\\
$t$ IRLS & 0.52 & 0.58 & 0.70 & 0.99 & 1.3 & 2.0\\
$t$-mixture ECM & 15.3 & 9.8 & 13.0 & 19.9 & 33.8 & 74.0\\
Logistic propensity & 1.6 & 2.0 & 2.2 & 2.2 & 3.7 & 4.6\\
Full pipeline (7 estimators, K=3) & 76.0 & 75.1 & 83.8 & 124 & 216 & 370\\
\midrule ECM iterations & 155 & 72 & 62 & 58 & 58 & 62\\
\bottomrule\end{tabular}\end{table}

\section{Application: drug package and coverage in the DRC}\label{sec:app}
\subsection{Question, sample and covariates}
We ask whether rounds that delivered praziquantel together with albendazole or mebendazole (``integrated'', $A=1$) achieved different coverage of school-age children than rounds that delivered praziquantel alone ($A=0$). The unit is a health-zone round: a zone and year in which a SAC round was recorded. There are \nApp{} such rounds in \nZones{} zones, \nInt{} of them integrated (\pctInt\%). We study three outcomes, all in the ESPEN file:
\begin{itemize}
\item \textbf{programme coverage}, $100\times$treated$/$targeted, the indicator programmes usually report;
\item \textbf{epidemiological coverage}, $100\times$treated$/$requiring, which uses the population in need as denominator;
\item the \textbf{targeting ratio}, $100\log($targeted$/$requiring$)$, which is not an outcome of interest but a diagnostic: delivery itself should not change how many people a plan counts as targeted relative to need.
\end{itemize}
Since $\log(\text{programme})=\log(\text{epidemiological})-\log(\text{targeted}/\text{requiring})$, any difference in programme coverage that is not matched by a difference in epidemiological coverage must come from the denominator.

The covariates, all fixed before the round, are the endemicity class (low, moderate, high), the log SAC population requiring PC, whether adults were also eligible (a marker of community-wide treatment), the number of earlier SAC rounds in the zone, whether a round took place the year before, calendar year, and indicators of the former provinces (\cref{tab:appsample}). The 26 current provinces are grouped into the former provinces to keep positivity; Kinshasa, with only four rounds, is grouped with Kongo-Central.

\needspace{14\baselineskip}
{\small\setlength{\LTcapwidth}{\textwidth}
\begin{longtable}{@{}>{\raggedright\arraybackslash}p{6.2cm}rrr@{}}
\caption{DRC analysis sample: rounds of preventive chemotherapy for school-age children, ESPEN health-zone data 2014--2025 ($n=\nApp$ rounds in $\nZones$ health zones). Binary covariates: percentage; others: mean (SD). Province indicators refer to the former provinces (reference: Bandundu). SMD: standardized mean difference.}\label{tab:appsample}\\
\toprule\rowcolor{rowshade}\textbf{Variable}&\textbf{PZQ alone}&\textbf{PZQ+ALB/MBD}&\textbf{SMD}\\\midrule\endfirsthead
\toprule\rowcolor{rowshade}\textbf{Variable}&\textbf{PZQ alone}&\textbf{PZQ+ALB/MBD}&\textbf{SMD}\\\midrule\endhead
Moderate endemicity & 62.5\% & 62.0\% & -0.01\\
High endemicity & 10.7\% & 11.8\% & +0.03\\
log SAC requiring PC & 10.39 (0.67) & 10.26 (0.65) & -0.21\\
Adults also eligible & 73.2\% & 73.8\% & +0.01\\
Previous rounds & 2.06 (1.91) & 2.35 (2.09) & +0.14\\
Round in previous year & 26.3\% & 32.8\% & +0.14\\
Calendar year & 1.42 (3.21) & 1.85 (3.23) & +0.13\\
Equateur & 12.5\% & 7.1\% & -0.18\\
Kasai-Occidental & 1.0\% & 4.9\% & +0.23\\
Kasai-Oriental & 13.6\% & 5.1\% & -0.29\\
Katanga & 22.3\% & 5.6\% & -0.49\\
Kinshasa/Kongo-Central & 8.2\% & 24.8\% & +0.46\\
Maniema & 2.9\% & 19.4\% & +0.54\\
Nord-Kivu & 4.8\% & 15.7\% & +0.36\\
Orientale & 25.5\% & 10.3\% & -0.41\\
Sud-Kivu & 5.2\% & 5.4\% & +0.01\\
\textit{Outcome:} Programme coverage (\%) & 54.6 (19.8) & 73.3 (26.3) & +0.81\\
\textit{Outcome:} Epidemiological coverage (\%) & 95.2 (12.6) & 96.9 (7.9) & +0.16\\
\textit{Outcome:} Targeting ratio (100 log) & 60.2 (48.7) & 35.2 (43.8) & -0.54\\
\bottomrule
\end{longtable}}

\subsection{Overlap and balance}
\begin{figure}[!htb]\centering\includegraphics[width=\textwidth]{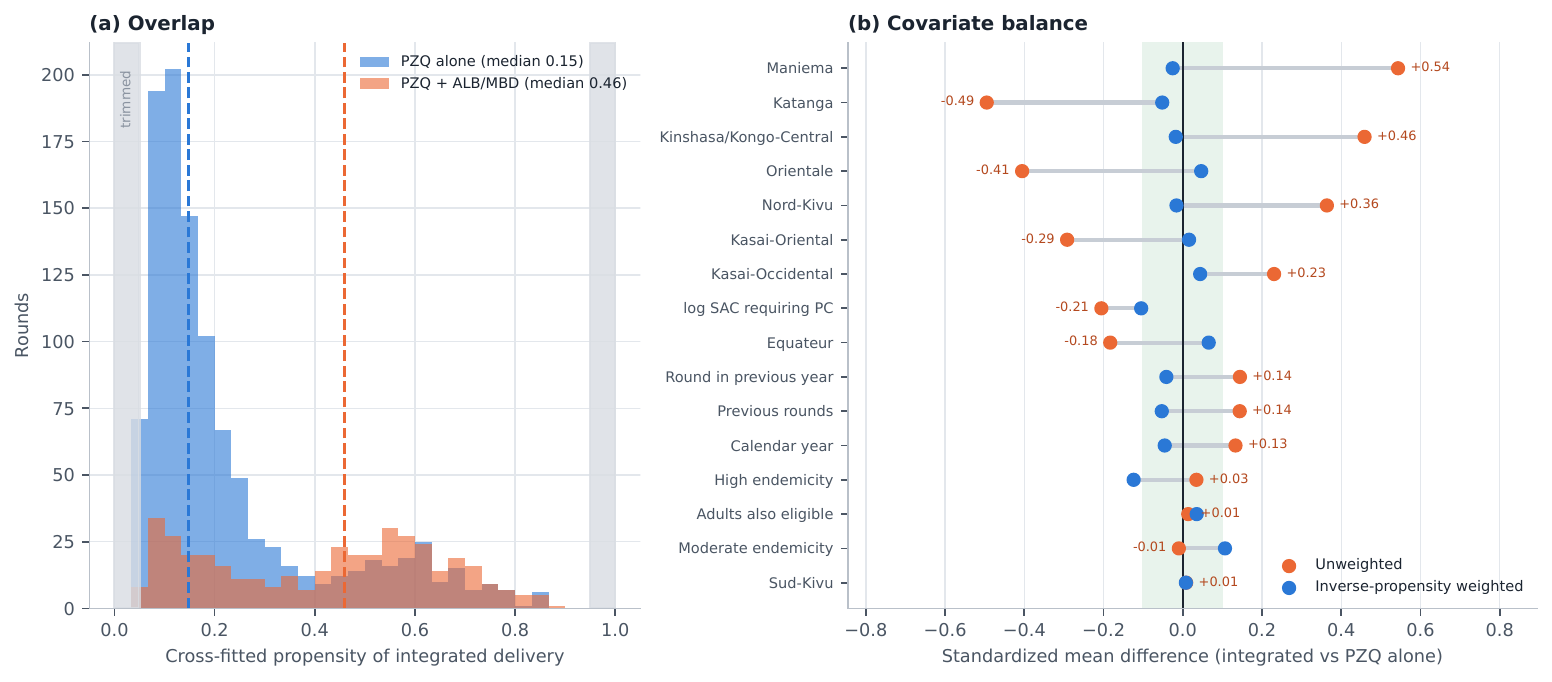}
\caption{Design diagnostics for the DRC application. (a) Cross-fitted propensity of integrated delivery by drug package; dashed lines are medians, grey bands the truncation region. (b) Standardized mean differences before (orange, labelled) and after (blue) inverse-propensity weighting; the green band marks $|{\rm SMD}|<0.1$.}\label{fig:appdesign}\end{figure}

The cross-fitted propensity of integrated delivery ranges from \psMin{} to \psMax, and \pctTrim\% of rounds reach the truncation bound (\cref{fig:appdesign}a). Weighting by the inverse propensity reduces the largest absolute standardized mean difference from \smdMaxBefore{} to \smdMaxAfter{} (\cref{fig:appdesign}b; \nSmdOver{} covariates remain above 0.1). Balance on measured covariates says nothing about unmeasured ones, such as the implementing partner, which plausibly determines both the drug package and the planning conventions.

\subsection{Residuals}
\begin{figure}[!htb]\centering\includegraphics[width=.92\textwidth]{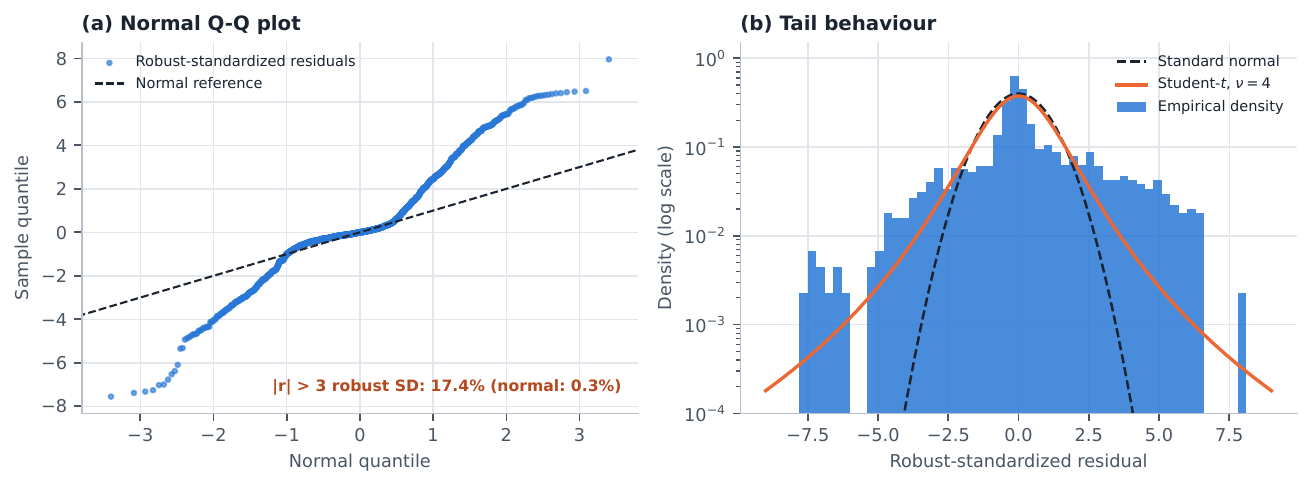}
\caption{Residuals of the arm-specific Student-$t$ fits for programme coverage in the DRC, standardized by a robust scale. (a) Normal Q--Q plot; (b) density on a log scale with normal and $t_4$ references.}\label{fig:appresid}\end{figure}

Residuals of the arm-specific $t$ fits for programme coverage are far from normal, though not in the textbook way (\cref{fig:appresid}). The heaps of \cref{fig:heaping} create a sharp central spike, so the robust scale is small and \appPctOut\% of the residuals lie more than three robust standard deviations from the median, against 0.3\% under normality; the excess kurtosis, by contrast, is only \appKurt. A spike plus broad shoulders is a shape for which a heavy-tailed working likelihood looks attractive, and also one for which a scale mixture may degenerate (\cref{sec:mc_em}).

\subsection{Estimates}
\begin{figure}[!htb]\centering\includegraphics[width=\textwidth]{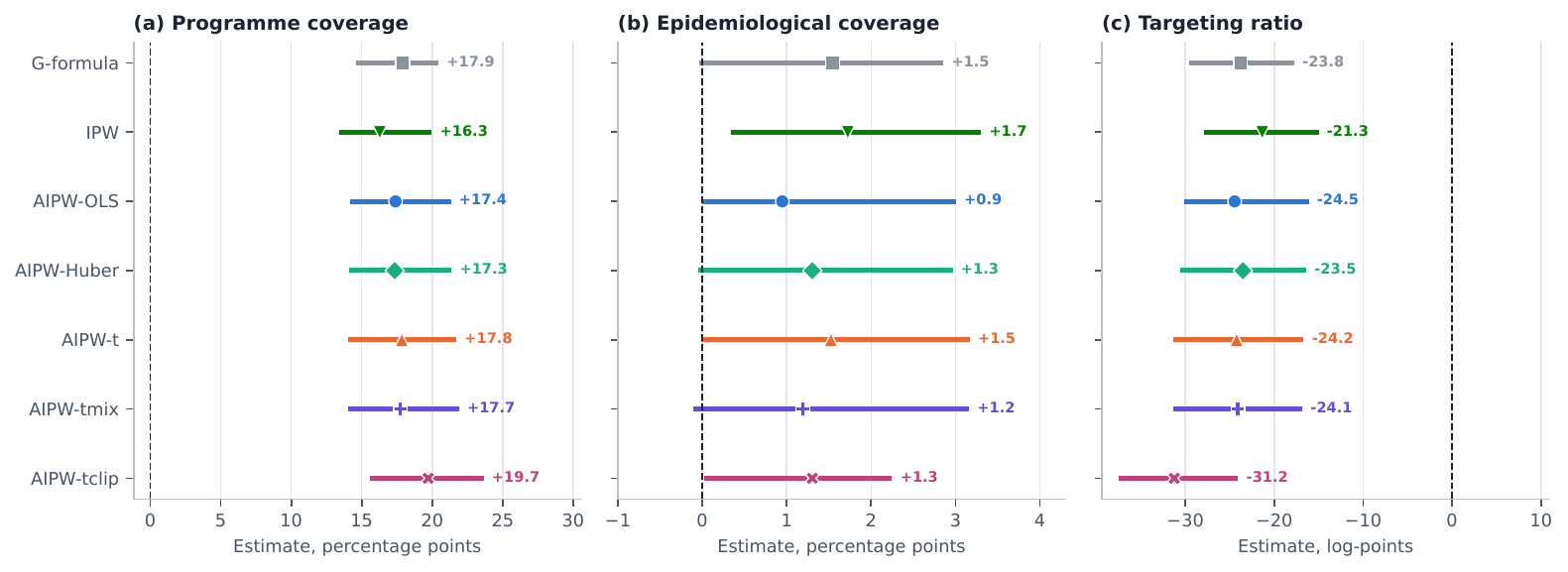}
\caption{Integrated delivery versus PZQ alone, DRC 2014--2025: point estimates (symbols, labelled) and 95\% cluster-bootstrap percentile intervals ($B=\Bboot$ resamples of health zones, whole pipeline repeated). (a) Programme coverage and (b) epidemiological coverage in percentage points; (c) targeting ratio in log-points ($100\log$).}\label{fig:appest}\end{figure}

\Cref{fig:appest,tab:appest} give the results. On \textbf{programme coverage}, integrated rounds score much higher: the four unclipped AIPW variants give between \appProgUncLo{} and \appProgUncHi{} percentage points (AIPW-OLS \appProgOls, cluster SE \appProgSeOls, bootstrap interval [\appProgLoOls, \appProgHiOls]), close to the unadjusted difference of \naiveProg. On \textbf{epidemiological coverage} the difference all but disappears: \appEpiUncLo{} to \appEpiUncHi{} points (AIPW-OLS \appEpiOls, interval [\appEpiLoOls, \appEpiHiOls]). The \textbf{targeting ratio} explains the gap: it is lower by \appDenAbsOls{} log-points in integrated rounds (AIPW-OLS; interval [\appDenLoOls, \appDenHiOls]), so that, for the same need, they targeted about \denPctOls\% fewer children. In other words, the advantage of about \appProgAbsOls{} points on the reported indicator is a difference in how the denominator was set, not in how many children were reached.

Three computational points stand out. First, the outcome engine barely matters: on every outcome the Huber, $t$ and mixture engines move the AIPW estimate by about one point or less, well inside its standard error, as \cref{prop:eff} and E1 lead one to expect. Second, clipping changes the estimand where residuals are skewed. On the targeting ratio AIPW-$t$clip gives \appDenClip, against \appDenT{} for AIPW-$t$, a shift of the kind \cref{prop:clip} and E6 predict, not a gain in precision. Third, clustering matters unevenly. For programme coverage the i.i.d.\ standard error of AIPW-OLS (\appProgSeIidOls) is even slightly larger than the cluster-robust one (\appProgSeOls); for epidemiological coverage it is smaller (\appEpiSeIidOls{} against \appEpiSeOls). The cluster bootstrap is the safer basis for intervals, because it also propagates the variability of the fold assignment and of the clipping scale.

\needspace{14\baselineskip}
{\small\setlength{\LTcapwidth}{\textwidth}
\begin{longtable}{@{}lrrrr@{}}
\caption{Integrated delivery (PZQ+ALB/MBD) versus PZQ alone in the DRC, three outcomes. SE (cl.): cluster-robust influence-function standard error with health zones as clusters; SE (iid): the same ignoring clustering; CI: cluster-bootstrap percentile interval ($B=\Bboot$ resamples of health zones, the whole cross-fitting pipeline repeated in each).}\label{tab:appest}\\
\toprule\rowcolor{rowshade}\textbf{Method}&\textbf{Estimate}&\textbf{SE (cl.)}&\textbf{SE (iid)}&\textbf{95\% bootstrap CI}\\\midrule\endfirsthead
\toprule\rowcolor{rowshade}\textbf{Method}&\textbf{Estimate}&\textbf{SE (cl.)}&\textbf{SE (iid)}&\textbf{95\% bootstrap CI}\\\midrule\endhead
\rowcolor{rowshade}\textbf{Programme coverage (percentage points)}&&&&\\
G-formula & +17.90 & -- & -- & [+14.60, +20.44]\\
IPW & +16.28 & -- & -- & [+13.40, +19.97]\\
AIPW-OLS & +17.40 & 1.71 & 1.88 & [+14.21, +21.32]\\
AIPW-Huber & +17.34 & 1.71 & 1.88 & [+14.13, +21.37]\\
AIPW-$t$ & +17.84 & 1.77 & 1.90 & [+14.07, +21.71]\\
AIPW-$t$mix & +17.75 & 1.78 & 1.90 & [+14.05, +21.90]\\
AIPW-$t$clip & +19.72 & 1.74 & 1.86 & [+15.60, +23.65]\\
\rowcolor{rowshade}\textbf{Epidemiological coverage (percentage points)}&&&&\\
G-formula & +1.54 & -- & -- & [-0.03, +2.86]\\
IPW & +1.72 & -- & -- & [+0.33, +3.30]\\
AIPW-OLS & +0.95 & 0.87 & 0.72 & [+0.00, +3.01]\\
AIPW-Huber & +1.30 & 0.81 & 0.76 & [-0.05, +2.97]\\
AIPW-$t$ & +1.52 & 0.90 & 0.78 & [-0.02, +3.17]\\
AIPW-$t$mix & +1.19 & 0.85 & 0.78 & [-0.11, +3.16]\\
AIPW-$t$clip & +1.30 & 0.41 & 0.32 & [+0.03, +2.25]\\
\rowcolor{rowshade}\textbf{Targeting ratio, 100 log(targeted/required)}&&&&\\
G-formula & -23.78 & -- & -- & [-29.58, -17.76]\\
IPW & -21.34 & -- & -- & [-27.94, -15.01]\\
AIPW-OLS & -24.45 & 3.35 & 3.30 & [-30.15, -16.12]\\
AIPW-Huber & -23.53 & 3.25 & 3.29 & [-30.63, -16.40]\\
AIPW-$t$ & -24.23 & 3.32 & 3.30 & [-31.41, -16.69]\\
AIPW-$t$mix & -24.09 & 3.36 & 3.32 & [-31.35, -16.84]\\
AIPW-$t$clip & -31.24 & 3.00 & 3.00 & [-37.49, -24.06]\\
\bottomrule
\end{longtable}}

\subsection{Influence and sensitivity}
\begin{figure}[!htb]\centering\includegraphics[width=\textwidth]{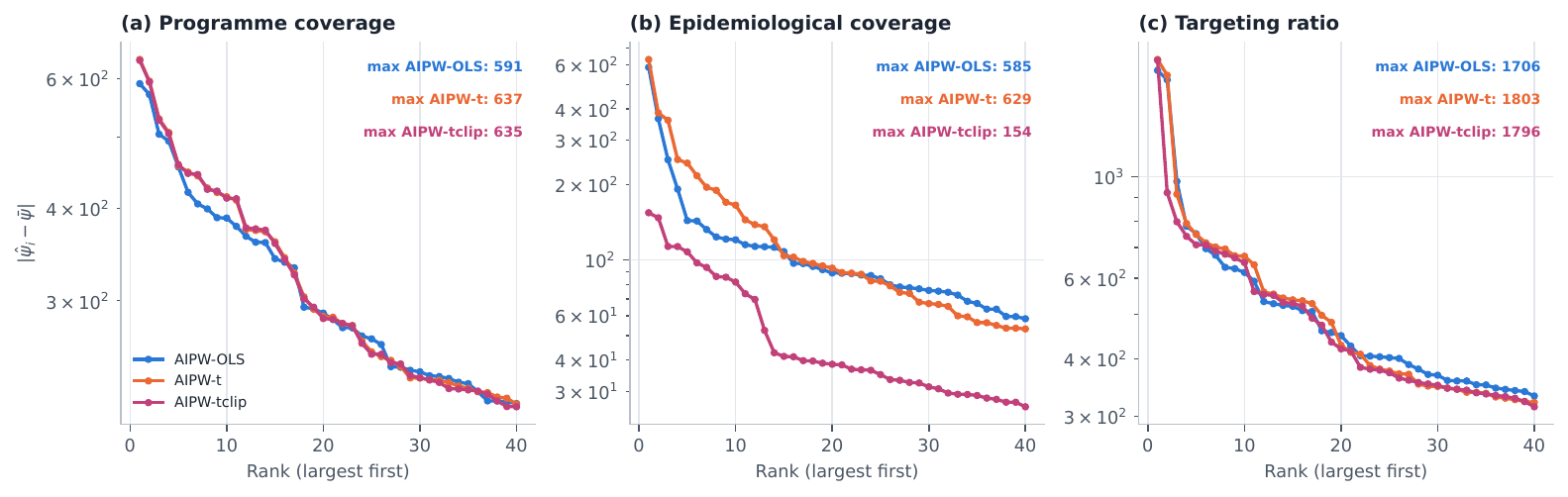}
\caption{Influence-function diagnostics on the DRC data: the forty largest absolute deviations $|\hat\psi_i-\bar\psi|$ for AIPW-OLS, AIPW-$t$ and AIPW-$t$clip, with the maximum printed. (a) Programme coverage, (b) epidemiological coverage, (c) targeting ratio.}\label{fig:appinf}\end{figure}

\begin{figure}[!htb]\centering\includegraphics[width=\textwidth]{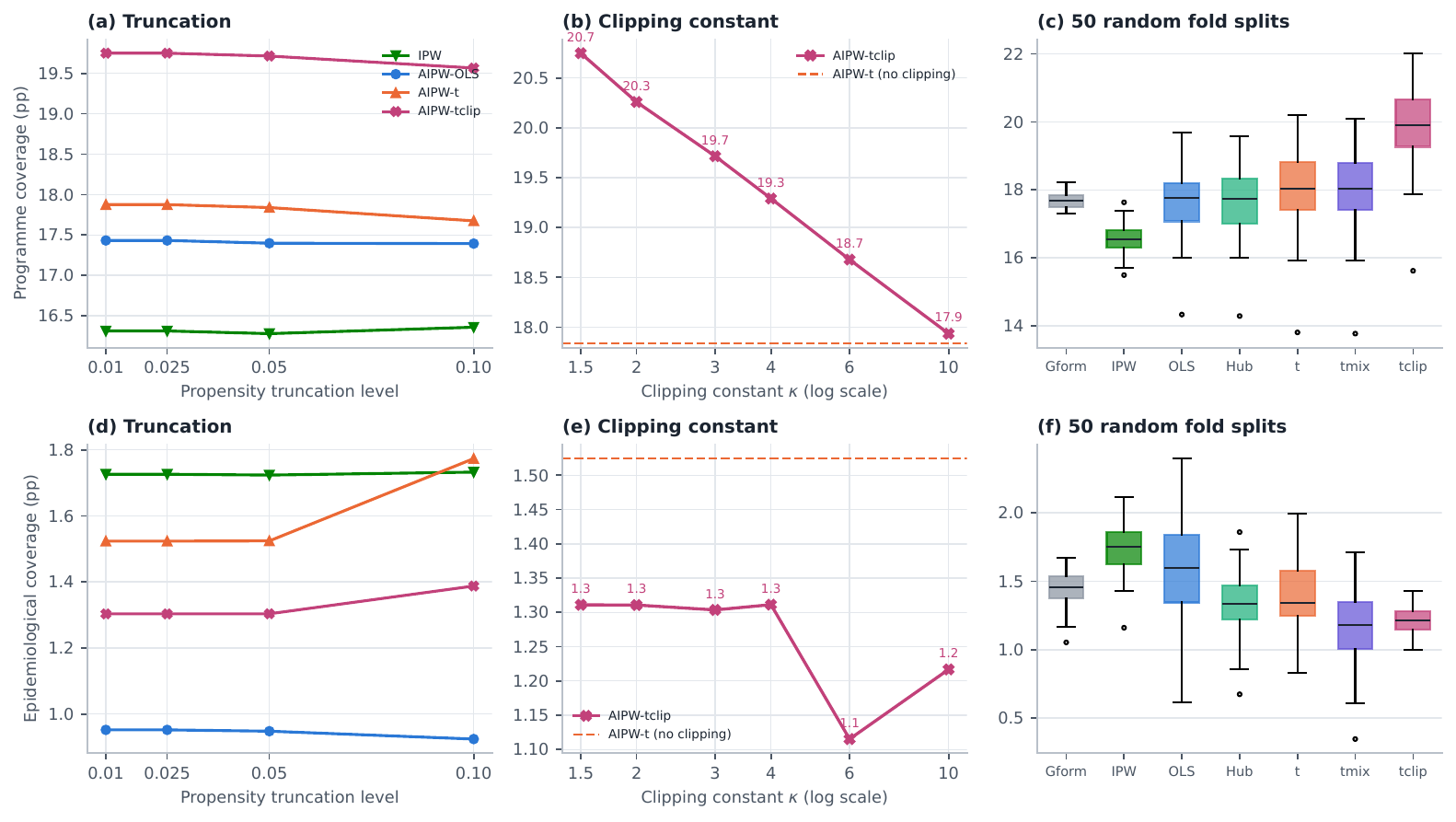}
\caption{Sensitivity of the DRC estimates; top row programme coverage, bottom row epidemiological coverage (percentage points). (a, d) Propensity truncation level. (b, e) Clipping constant $\kappa$ for AIPW-$t$clip, with AIPW-$t$ as reference (dashed). (c, f) Distribution of the estimates over \nSplits{} random zone-level fold assignments.}\label{fig:appsens}\end{figure}

The influence-function diagnostics (\cref{fig:appinf}) show where clipping acts and where it does not. For epidemiological coverage the largest absolute score deviation falls from \maxIfEpiOls{} (AIPW-OLS) to \maxIfEpiClip{} after clipping. For programme coverage it does not fall at all (\maxIfProgOls{} against \maxIfProgClip): the largest scores there come from rounds with small estimated propensities, whose weights clipping leaves untouched, rather than from extreme residuals. The estimates are insensitive to the propensity truncation level (AIPW-OLS between \trimProgLo{} and \trimProgHi{} points on programme coverage for truncation at 0.01 to 0.10; \cref{fig:appsens}a), while the clipped estimate drifts with $\kappa$ from \kappaProgHi{} to \kappaProgLo{} points as clipping becomes milder (\cref{fig:appsens}b). The largest source of variation is the fold assignment itself: over \nSplits{} random zone-level splits, AIPW-OLS ranges from \splitProgLo{} to \splitProgHi{} points on programme coverage and from \splitEpiLo{} to \splitEpiHi{} on epidemiological coverage (\cref{fig:appsens}c,f). With a few hundred clusters and five folds, a single cross-fit is a noisy summary, and reporting the median over repeated splits, as \citet{chernozhukov2018} recommend, is worth its modest cost. None of these checks changes the substantive conclusion.

\subsection{What the contrast means}
Under \cref{ass:consistency,ass:exch,ass:pos,ass:sutva} the epidemiological-coverage estimate would be the effect of the drug package on the share of children in need who were recorded as treated. We do not claim that interpretation. The drug package is chosen together with partners, funding and logistics that are not recorded in the file, the treated counts are capped at the requiring population, and the denominators are planning numbers. What the analysis does show is computational and methodological: a confounding-adjusted, doubly robust, cluster-aware comparison of a routinely reported indicator can be dominated by a definitional artefact that differs between arms. As E5 showed in simulation, no choice of outcome engine repairs this. The remedy is to choose the outcome, here epidemiological rather than programme coverage, by looking at how each indicator is produced.

\section{A target-trial template for effects on disease}\label{sec:drc}
The application concerns delivery, not health. Estimating the effect of PC strategies on infection in the DRC would need data that the public files do not contain. Following the target-trial logic of \citet{hernan2016,hernan2020}, \cref{tab:protocol} records what such a study would require. It is a design template, not a description of an existing dataset.
\begin{longtable}{@{}>{\raggedright\arraybackslash\bfseries\color{brand}}p{3.0cm}>{\raggedright\arraybackslash}p{12.2cm}@{}}
\caption{Target-trial template for evaluating preventive chemotherapy against schistosomiasis with routine data.}\label{tab:protocol}\\
\toprule\rowcolor{rowshade}\textbf{Element}&\textbf{Specification (to be agreed with programme and data owners)}\\\midrule\endfirsthead
\toprule\rowcolor{rowshade}\textbf{Element}&\textbf{Specification (continued)}\\\midrule\endhead
Unit&Health zone, school catchment or village cluster rather than the individual, because community treatment creates spill-overs.\\
Eligibility&Endemic units with a baseline parasitological survey, a stable catchment and adequate reporting completeness.\\
Time zero&Date at which eligibility and pre-treatment covariates are recorded.\\
Strategies&Well-defined programme intensities (e.g.\ annual versus biennial rounds, or on-time versus delayed rounds), only where ethically and operationally meaningful.\\
Outcome&Pre-specified infection prevalence or intensity at follow-up, from impact surveys rather than from delivery records.\\
Follow-up&Common observation window; explicit handling of migration, denominators and loss to follow-up.\\
Confounders&Baseline endemicity, WASH access, age structure, treatment history, implementing partner, geography, health-system access, socio-economic context.\\
Estimand&Zone-level ATE or policy contrast, interference-aware where needed.\\
Identification&Exchangeability, positivity, consistency and explicit spatial spill-over assumptions.\\
Analysis&Cross-fitting with zone-level folds; outcome engine and clipping chosen by pre-specified diagnostics (\cref{sec:discussion}); cluster-robust and cluster-bootstrap inference; median over repeated splits.\\
Sensitivity&Unmeasured confounding, propensity truncation, alternative tail models, artefact structure (\cref{fig:artefacts}), denominators and capping (\cref{fig:denominator}), negative controls.\\\bottomrule
\end{longtable}
\paragraph{Why delivery records are not enough.} Programme totals cannot identify an effect on infection: coverage is confounded by targeting, secular trends and denominator revisions, and, as \cref{sec:app} shows, coverage itself can be an artefact of planning conventions. The ESPEN and WHO files are best used to define sampling frames, choose comparison units and negotiate access to linked parasitological data \citep{who_pct,espen}.

\section{Discussion}\label{sec:discussion}
\begin{keybox}[title={\sffamily\bfseries\small Take-home messages},fonttitle=\sffamily\bfseries\small,coltitle=green!35!black,colbacktitle=softgreen,attach title to upper={\ }]
\small
\begin{enumerate}[leftmargin=1.2em,itemsep=1pt]
\item Making only the outcome \emph{regression} robust (Huber, $t$, $t$-mixture) leaves cross-fitted AIPW unchanged in bias, RMSE, coverage and influence function, in simulation and on the DRC data (\cref{prop:eff}).
\item Bounding the \emph{augmentation residual} removes much of the variance due to heavy tails and symmetric artefacts, but it changes the estimand when residuals are skewed differently in the two arms (\cref{prop:clip}, E6, DRC targeting ratio) and its intervals need a better variance estimator.
\item Differential recording artefacts, such as arm-specific denominators, bias every method. They are a data-production problem, not a regression problem.
\item Heaped data break naive implementations: guard the MAD against collapse and bound mixture scales away from zero.
\end{enumerate}
\end{keybox}
\paragraph{Recommendations for practice.} When outcomes are heavy-tailed, heaped or capped, we suggest (i) inspecting the influence function before choosing an estimator (\cref{fig:appinf}); (ii) reporting standard AIPW together with a bounded-influence variant and explaining any difference in terms of the estimand; (iii) checking each component of a ratio outcome separately, as the targeting ratio did here; (iv) using zone-level folds, cluster-robust standard errors, a cluster bootstrap and the median over repeated splits; and (v) investigating how artefacts arise in each arm before trusting any adjustment.

\paragraph{Limitations.} (i) The simulation model has $p=3$ covariates, linear working models and independent units. (ii) The misspecification design is mild and cannot separate estimators on bias. (iii) The $t$-mixture shares its location across components and fixes $\nu=4$; it is an ECM algorithm, not the variational mixture of experts of \cref{app:vb}. (iv) The clipping constant $\kappa=3$ was fixed in advance, and the clipped standard error ignores estimation of the clipping scale. (v) Propensity truncation changes the target population. (vi) In the application, unmeasured confounding (partners, funding) is likely, treated counts are capped, and residuals are resampled independently in the plasmode experiment, which ignores within-zone dependence. (vii) We did not search the literature systematically for winsorized or bounded-influence AIPW estimators and claim no novelty for the clipped variant.

\paragraph{Future work.} A variance estimator for clipped AIPW that accounts for the clipping step; stronger misspecification and higher-dimensional covariates; estimation of $\nu$ and covariate-dependent gating in a variational mixture of experts, with its effect on the causal score; plasmode designs that resample whole zones; and a study of PC effects on infection along the lines of \cref{tab:protocol} once linked survey data are available.

\section{Conclusion}
Replacing least squares by Huber, Student-$t$ or Student-$t$-mixture outcome models did not improve cross-fitted AIPW, neither in a Monte Carlo study with up to \Breps{} replications per cell nor on schistosomiasis programme data from the DRC, exactly as the efficiency bound predicts. Clipping the augmentation term did help, substantially, for symmetric artefacts and heavy tails, but it targets a clipped-location contrast and needs better variance estimation. Differential artefacts defeated every estimator, and the DRC data contain a clear one: the apparent coverage advantage of integrated delivery is a difference in targeting denominators. The code that produces every number, table and figure in this paper is a single Python file.

\section*{Declarations}
\small
\textbf{Authorship.} Cosmas Kambale Musambi designed the study, assembled the data, wrote the code and drafted the manuscript. Ernest Fokou\'e contributed to the statistical framing, reviewed the methods and the manuscript, and approved its submission.
\textbf{Data and code availability.} The four public data files, the Python code and all intermediate results are included in the accompanying archive; the code is also provided as an ancillary file of this arXiv submission (\cref{app:repro}). The ESPEN and WHO data are subject to the terms of use of their providers.
\textbf{Ethics.} Only aggregate, publicly available programme counts are used; no individual is identifiable and no ethical approval was required.
\textbf{Funding.} This work received no specific funding.
\textbf{Competing interests.} The authors declare no competing interests.
\textbf{Use of AI tools.} An AI assistant (Claude, Anthropic) was used for code refactoring, typesetting and language editing. The authors checked the derivations, code, analyses and references and take full responsibility for the content.
\normalsize

\appendix
\section{Reproducibility}\label{app:repro}
The archive contains:
\begin{itemize}[itemsep=0pt]
\item \code{cosmas\_fokoue\_statco.py}, the single source file (analyses, simulations, figures, tables);
\item \code{data/raw/}, the four public data files, and \code{results/}, all simulation and analysis output;
\item \code{main.tex}, \code{abstract.tex}, \code{references.bib} and the compiled \code{main.pdf};
\item \code{figures/}, the 26 vector figures (the workflow diagram is drawn in TikZ), and \code{tables/}, the tables and number macros.
\end{itemize}
No number in the text, tables or figures is typed by hand: all are written by the code into \code{tables/numbers.tex}.
\begin{tcolorbox}[colback=rowshade,colframe=ink2,boxrule=0pt,leftrule=3pt,arc=0pt,fontupper=\ttfamily\small]
pip install -r requirements.txt\\
python cosmas\_fokoue\_statco.py all\ \ \ \ \ \ \ \ \ \# everything (about 40 min, one core)\\
python cosmas\_fokoue\_statco.py figures\ \ \ \ \ \# figures and tables from stored results\\
python cosmas\_fokoue\_statco.py quick\ \ \ \ \ \ \ \# smoke test (under one minute)\\
pdflatex main \&\& bibtex main \&\& pdflatex main \&\& pdflatex main
\end{tcolorbox}
Software: Python~3 with NumPy, pandas, SciPy, scikit-learn and Matplotlib \citep{harris2020,virtanen2020,pedregosa2011,hunter2007}; versions are listed in \code{requirements.txt}. All random draws use fixed seeds, and the fold assignments are fixed by the replication index. The results are not intended for clinical or programme decisions.

\section{Proofs}\label{app:proofs}
\begin{proofof}{\cref{prop:bias}}
Condition on $X$. By consistency and exchangeability, $\E\{A(Y-\bar m_1)\mid X\}=e(X)\{m_1(X)-\bar m_1(X)\}$ and $\E\{(1-A)(Y-\bar m_0)\mid X\}=\{1-e(X)\}\{m_0(X)-\bar m_0(X)\}$. Hence
\[\E\{\psi\mid X\}=\bar m_1-\bar m_0+\frac{e}{\bar e}(m_1-\bar m_1)-\frac{1-e}{1-\bar e}(m_0-\bar m_0).\]
Subtracting $\tau(X)=m_1-m_0$ and collecting terms,
\[\begin{aligned}\E\{\psi\mid X\}-\tau(X)&=(\bar m_1-m_1)\Big(1-\frac{e}{\bar e}\Big)-(\bar m_0-m_0)\Big(1-\frac{1-e}{1-\bar e}\Big)\\&=(\bar e-e)\Big\{\frac{\bar m_1-m_1}{\bar e}+\frac{\bar m_0-m_0}{1-\bar e}\Big\}.\end{aligned}\]
Taking expectations over $X$ gives the claim.
\end{proofof}
\begin{proofof}{\cref{prop:eff}}
The efficient influence function is $\varphi=\tau(X)-\ATE+\frac{A}{e}\varepsilon_1-\frac{1-A}{1-e}\varepsilon_0$ with $\varepsilon_a=Y-m_a(X)$ \citep{hahn1998,tsiatis2006}. The three terms are uncorrelated (the residual terms have conditional mean zero and $A(1-A)=0$), so $\E\varphi^2=\Var\tau(X)+\E\{A\varepsilon_1^2/e^2\}+\E\{(1-A)\varepsilon_0^2/(1-e)^2\}=\Var\tau(X)+\E\{\sigma_1^2/e+\sigma_0^2/(1-e)\}=V$. Asymptotic linearity of the cross-fitted estimator with influence function $\varphi$ under the stated rate condition is Theorem~5.1 of \citet{chernozhukov2018}. Since $\varphi$ does not involve the nuisance estimator, the limit variance is the same for every consistent engine.
\end{proofof}
\begin{proofof}{\cref{prop:clip}}
With $\bar e=e$, $\E\{\tfrac{A}{e}c(Y-\bar m_1)\mid X\}=\E\{c(Y(1)-\bar m_1)\mid A=1,X\}=\delta_1(X)$ by exchangeability, and similarly for the control term. The first display follows. If $Y(a)-\bar m_a(X)$ is symmetric about $0$ given $X$, $c$ odd implies $\delta_a=0$. If $\bar m_a=\ell_a$ solves $\E\{c(Y-\ell_a)\mid A=a,X\}=0$, then $\delta_a=0$ by definition.
\end{proofof}
\begin{proofof}{\cref{prop:em}}
The complete-data log-likelihood is linear in the sufficient statistics $r_{ik}$, $u_{ik}$, $r_{ik}u_{ik}$, $r_{ik}u_{ik}y_i$, $r_{ik}u_{ik}y_i^2$ of the exponential-family representation \eqref{eq:mix}. The E-step \eqref{eq:estep} computes their exact conditional expectations: the posterior of $U_i$ given $y_i$ and $Z_i=k$ is $\mathrm{Gamma}\{(\nu+1)/2,\ (\nu+(y_i-\beta^\top D_i)^2/\sigma_k^2)/2\}$ with mean $u_{ik}$. Each conditional-maximization step maximizes the expected complete-data log-likelihood $Q$ over one block of parameters ($\beta$; then $\sigma,\pi$) given the others. For $\sigma_k^2$ the relevant part of $Q$ is $-\tfrac12 n_k\log\sigma_k^2-\tfrac12 S_k/\sigma_k^2$ with $n_k=\sum_ir_{ik}$ and $S_k=\sum_ir_{ik}u_{ik}(y_i-\beta^\top D_i)^2$, which is unimodal in $\sigma_k^2$ with maximum at $S_k/n_k$; its maximum over $\sigma_k^2\ge\underline\sigma^2$ is therefore attained at $\max(S_k/n_k,\underline\sigma^2)$. No step can decrease $Q$, and the EM inequality $\ell(\theta^{(r+1)})-\ell(\theta^{(r)})\ge Q(\theta^{(r+1)}\mid\theta^{(r)})-Q(\theta^{(r)}\mid\theta^{(r)})\ge0$ gives the claim \citep{dempster1977,meng1993}.
\end{proofof}

\section{Algorithm summary}\label{app:algo}
\begin{tcolorbox}[colback=white,colframe=brand,title={\sffamily\bfseries Cross-fitted AIPW with a chosen outcome engine},fonttitle=\sffamily\bfseries\small]
\small\begin{enumerate}[leftmargin=1.4em,itemsep=1pt]
\item Split the sample into $K$ folds (seeded); with clustered data, assign whole clusters to folds.
\item For each fold $k$: fit a logistic propensity model on the other folds and truncate its predictions to $[0.05,0.95]$.
\item For each fold $k$ and arm $a$: fit $\hat m_a$ on the arm-$a$ training units with the chosen engine (OLS, Huber, $t$-IRLS or $t$-mixture ECM).
\item For held-out units: predict both counterfactual means and form the score \eqref{eq:aipw}, clipping the residuals at $\pm3\hat s_a$ for the clipped variant (MAD scale with the collapse guard of \cref{sec:numerics}).
\item Return $\hat\tau=\bar\psi$, $\widehat{\mathrm{SE}}=\mathrm{sd}(\psi)/\sqrt n$ and, with clusters, $\widehat{\mathrm{SE}}_{\rm cl}$ of \eqref{eq:secl}; for intervals, repeat steps 1--5 on cluster-bootstrap resamples.
\end{enumerate}
\end{tcolorbox}
\begin{tcolorbox}[colback=white,colframe=brand,title={\sffamily\bfseries ECM for the two-component Student-$t$ scale-mixture regression},fonttitle=\sffamily\bfseries\small]
\small\begin{enumerate}[leftmargin=1.4em,itemsep=1pt]
\item Initialize $\beta$ by OLS, $\sigma=s_0\cdot(0.5,2.0)$ with $s_0$ the MAD of the OLS residuals, $\pi=(\tfrac12,\tfrac12)$; set $\underline\sigma=10^{-3}s_0$.
\item \textbf{E-step:} compute $r_{ik}$ and $u_{ik}$ from \eqref{eq:estep}; record the observed-data log-likelihood.
\item \textbf{CM-1:} $\beta\leftarrow(D^\top WD)^{-1}D^\top Wy$, $w_i=\sum_kr_{ik}u_{ik}/\sigma_k^2$ (least-squares fallback if singular).
\item \textbf{CM-2:} $\sigma_k^2\leftarrow\max\{\sum_ir_{ik}u_{ik}(y_i-\beta^\top D_i)^2/\sum_ir_{ik},\ \underline\sigma^2\}$, $\pi_k\leftarrow n^{-1}\sum_ir_{ik}$.
\item Stop when the relative log-likelihood change is below $10^{-9}$ (at most 300 iterations).
\end{enumerate}
\end{tcolorbox}

\section{Beyond EM: a variational mixture of experts (not implemented)}\label{app:vb}
The natural extension replaces the shared linear location by experts $\mu_k(a,x)$ with a gating network $\pi_k(a,x)$ and places priors on the parameters. With scalar outcome and $q(Z,U)=q(Z)q(U\mid Z)$, the evidence lower bound is
\[\mathcal L(q,\Theta)=\E_q\log p(Y,Z,U\mid X,A;\Theta)-\E_q\log q(Z,U),\]
the optimal $q(U\mid Z=k)$ is $\mathrm{Gamma}\{(\nu_k+1)/2,(\nu_k+\delta_{ik})/2\}$ with $\delta_{ik}=(y_i-\mu_k)^2/\sigma_k^2$, and $q(Z)$ is proportional to the exponentiated expected complete-data log-likelihood. Updates for $\nu_k$ and the gating parameters, convergence guarantees, and the effect of approximate inference on the causal score all need separate derivation and testing; the degeneracy seen on heaped DRC data (\cref{sec:mc_em}) suggests that priors on the scales would be essential. Student-$t$ mixtures \citep{peel2000}, mixtures of experts \citep{jordan1994}, variational methods \citep{blei2017,bishop2006} and doubly robust estimation \citep{bang2005} are each well studied, so a literature search must precede any novelty claim.

\bibliographystyle{plainnat}
\bibliography{references}

\begin{thebibliography}{38}
\providecommand{\natexlab}[1]{#1}
\providecommand{\url}[1]{\texttt{#1}}
\expandafter\ifx\csname urlstyle\endcsname\relax
  \providecommand{\doi}[1]{doi: #1}\else
  \providecommand{\doi}{doi: \begingroup \urlstyle{rm}\Url}\fi

\bibitem[Bang and Robins(2005)]{bang2005}
Heejung Bang and James~M. Robins.
\newblock Doubly robust estimation in missing data and causal inference models.
\newblock \emph{Biometrics}, 61\penalty0 (4):\penalty0 962--973, 2005.
\newblock URL \url{https://doi.org/10.1111/j.1541-0420.2005.00377.x}.

\bibitem[Bishop(2006)]{bishop2006}
Christopher~M. Bishop.
\newblock \emph{Pattern Recognition and Machine Learning}.
\newblock Springer, New York, 2006.

\bibitem[Blei et~al.(2017)Blei, Kucukelbir, and McAuliffe]{blei2017}
David~M. Blei, Alp Kucukelbir, and Jon~D. McAuliffe.
\newblock Variational inference: a review for statisticians.
\newblock \emph{Journal of the American Statistical Association}, 112\penalty0
  (518):\penalty0 859--877, 2017.
\newblock URL \url{https://doi.org/10.1080/01621459.2017.1285773}.

\bibitem[Cameron and Miller(2015)]{cameron2015}
A.~Colin Cameron and Douglas~L. Miller.
\newblock A practitioner's guide to cluster-robust inference.
\newblock \emph{Journal of Human Resources}, 50\penalty0 (2):\penalty0
  317--372, 2015.
\newblock URL \url{https://doi.org/10.3368/jhr.50.2.317}.

\bibitem[Chernozhukov et~al.(2018)Chernozhukov, Chetverikov, Demirer, Duflo,
  Hansen, Newey, and Robins]{chernozhukov2018}
Victor Chernozhukov, Denis Chetverikov, Mert Demirer, Esther Duflo, Christian
  Hansen, Whitney Newey, and James Robins.
\newblock Double/debiased machine learning for treatment and structural
  parameters.
\newblock \emph{The Econometrics Journal}, 21\penalty0 (1):\penalty0 C1--C68,
  2018.
\newblock URL \url{https://doi.org/10.1111/ectj.12097}.

\bibitem[Chiang et~al.(2022)Chiang, Kato, Ma, and Sasaki]{chiang2022}
Harold~D. Chiang, Kengo Kato, Yukun Ma, and Yuya Sasaki.
\newblock Multiway cluster robust double/debiased machine learning.
\newblock \emph{Journal of Business \& Economic Statistics}, 40\penalty0
  (3):\penalty0 1046--1056, 2022.
\newblock URL \url{https://doi.org/10.1080/07350015.2021.1895815}.

\bibitem[Colley et~al.(2014)Colley, Bustinduy, Secor, and King]{colley2014}
Daniel~G. Colley, Amaya~L. Bustinduy, W.~Evan Secor, and Charles~H. King.
\newblock Human schistosomiasis.
\newblock \emph{The Lancet}, 383\penalty0 (9936):\penalty0 2253--2264, 2014.
\newblock URL \url{https://doi.org/10.1016/S0140-6736(13)61949-2}.

\bibitem[Crump et~al.(2009)Crump, Hotz, Imbens, and Mitnik]{crump2009}
Richard~K. Crump, V.~Joseph Hotz, Guido~W. Imbens, and Oscar~A. Mitnik.
\newblock Dealing with limited overlap in estimation of average treatment
  effects.
\newblock \emph{Biometrika}, 96\penalty0 (1):\penalty0 187--199, 2009.
\newblock URL \url{https://doi.org/10.1093/biomet/asn055}.

\bibitem[Dempster et~al.(1977)Dempster, Laird, and Rubin]{dempster1977}
Arthur~P. Dempster, Nan~M. Laird, and Donald~B. Rubin.
\newblock Maximum likelihood from incomplete data via the {EM} algorithm.
\newblock \emph{Journal of the Royal Statistical Society: Series B},
  39\penalty0 (1):\penalty0 1--38, 1977.
\newblock URL \url{https://doi.org/10.1111/j.2517-6161.1977.tb01600.x}.

\bibitem[Franklin et~al.(2014)Franklin, Schneeweiss, Polinski, and
  Rassen]{franklin2014}
Jessica~M. Franklin, Sebastian Schneeweiss, Jennifer~M. Polinski, and Jeremy~A.
  Rassen.
\newblock Plasmode simulation for the evaluation of pharmacoepidemiologic
  methods in complex healthcare databases.
\newblock \emph{Computational Statistics \& Data Analysis}, 72:\penalty0
  219--226, 2014.
\newblock URL \url{https://doi.org/10.1016/j.csda.2013.10.018}.

\bibitem[Hahn(1998)]{hahn1998}
Jinyong Hahn.
\newblock On the role of the propensity score in efficient semiparametric
  estimation of average treatment effects.
\newblock \emph{Econometrica}, 66\penalty0 (2):\penalty0 315--331, 1998.
\newblock URL \url{https://doi.org/10.2307/2998560}.

\bibitem[Hampel et~al.(1986)Hampel, Ronchetti, Rousseeuw, and
  Stahel]{hampel1986}
Frank~R. Hampel, Elvezio~M. Ronchetti, Peter~J. Rousseeuw, and Werner~A.
  Stahel.
\newblock \emph{Robust Statistics: The Approach Based on Influence Functions}.
\newblock Wiley, New York, 1986.

\bibitem[Harris et~al.(2020)Harris, Millman, van~der Walt, et~al.]{harris2020}
Charles~R. Harris, K.~Jarrod Millman, St{\'e}fan~J. van~der Walt, et~al.
\newblock Array programming with {NumPy}.
\newblock \emph{Nature}, 585:\penalty0 357--362, 2020.
\newblock URL \url{https://doi.org/10.1038/s41586-020-2649-2}.

\bibitem[Heitjan and Rubin(1991)]{heitjan1991}
Daniel~F. Heitjan and Donald~B. Rubin.
\newblock Ignorability and coarse data.
\newblock \emph{The Annals of Statistics}, 19\penalty0 (4):\penalty0
  2244--2253, 1991.
\newblock URL \url{https://doi.org/10.1214/aos/1176348396}.

\bibitem[Hern{\'a}n and Robins(2016)]{hernan2016}
Miguel~A. Hern{\'a}n and James~M. Robins.
\newblock Using big data to emulate a target trial when a randomized trial is
  not available.
\newblock \emph{American Journal of Epidemiology}, 183\penalty0 (8):\penalty0
  758--764, 2016.
\newblock URL \url{https://doi.org/10.1093/aje/kwv254}.

\bibitem[Hern{\'a}n and Robins(2020)]{hernan2020}
Miguel~A. Hern{\'a}n and James~M. Robins.
\newblock \emph{Causal Inference: What If}.
\newblock Chapman \& Hall/CRC, Boca Raton, 2020.
\newblock URL
  \url{https://www.hsph.harvard.edu/miguel-hernan/causal-inference-book/}.

\bibitem[Huber(1964)]{huber1964}
Peter~J. Huber.
\newblock Robust estimation of a location parameter.
\newblock \emph{The Annals of Mathematical Statistics}, 35\penalty0
  (1):\penalty0 73--101, 1964.
\newblock URL \url{https://doi.org/10.1214/aoms/1177703732}.

\bibitem[Hudgens and Halloran(2008)]{hudgens2008}
Michael~G. Hudgens and M.~Elizabeth Halloran.
\newblock Toward causal inference with interference.
\newblock \emph{Journal of the American Statistical Association}, 103\penalty0
  (482):\penalty0 832--842, 2008.
\newblock URL \url{https://doi.org/10.1198/016214508000000292}.

\bibitem[Hunter(2007)]{hunter2007}
John~D. Hunter.
\newblock Matplotlib: a {2D} graphics environment.
\newblock \emph{Computing in Science \& Engineering}, 9\penalty0 (3):\penalty0
  90--95, 2007.
\newblock URL \url{https://doi.org/10.1109/MCSE.2007.55}.

\bibitem[Jordan and Jacobs(1994)]{jordan1994}
Michael~I. Jordan and Robert~A. Jacobs.
\newblock Hierarchical mixtures of experts and the {EM} algorithm.
\newblock \emph{Neural Computation}, 6\penalty0 (2):\penalty0 181--214, 1994.
\newblock URL \url{https://doi.org/10.1162/neco.1994.6.2.181}.

\bibitem[Kang and Schafer(2007)]{kang2007}
Joseph D.~Y. Kang and Joseph~L. Schafer.
\newblock Demystifying double robustness: a comparison of alternative
  strategies for estimating a population mean from incomplete data.
\newblock \emph{Statistical Science}, 22\penalty0 (4):\penalty0 523--539, 2007.
\newblock URL \url{https://doi.org/10.1214/07-STS227}.

\bibitem[Lange et~al.(1989)Lange, Little, and Taylor]{lange1989}
Kenneth~L. Lange, Roderick J.~A. Little, and Jeremy M.~G. Taylor.
\newblock Robust statistical modeling using the $t$ distribution.
\newblock \emph{Journal of the American Statistical Association}, 84\penalty0
  (408):\penalty0 881--896, 1989.
\newblock URL \url{https://doi.org/10.1080/01621459.1989.10478852}.

\bibitem[Liang and Zeger(1986)]{liang1986}
Kung-Yee Liang and Scott~L. Zeger.
\newblock Longitudinal data analysis using generalized linear models.
\newblock \emph{Biometrika}, 73\penalty0 (1):\penalty0 13--22, 1986.
\newblock URL \url{https://doi.org/10.1093/biomet/73.1.13}.

\bibitem[Meng and Rubin(1993)]{meng1993}
Xiao-Li Meng and Donald~B. Rubin.
\newblock Maximum likelihood estimation via the {ECM} algorithm: a general
  framework.
\newblock \emph{Biometrika}, 80\penalty0 (2):\penalty0 267--278, 1993.
\newblock URL \url{https://doi.org/10.1093/biomet/80.2.267}.

\bibitem[Morris et~al.(2019)Morris, White, and Crowther]{morris2019}
Tim~P. Morris, Ian~R. White, and Michael~J. Crowther.
\newblock Using simulation studies to evaluate statistical methods.
\newblock \emph{Statistics in Medicine}, 38\penalty0 (11):\penalty0 2074--2102,
  2019.
\newblock URL \url{https://doi.org/10.1002/sim.8086}.

\bibitem[{Our World in Data}(2026{\natexlab{a}})]{owid_cov}
{Our World in Data}.
\newblock Schistosomiasis treatment coverage (data adapted from the {WHO}
  global health observatory), 2026{\natexlab{a}}.
\newblock URL
  \url{https://ourworldindata.org/grapher/schistosomiasis-treatment-coverage}.
\newblock Data downloaded 13 November 2026.

\bibitem[{Our World in Data}(2026{\natexlab{b}})]{owid_req}
{Our World in Data}.
\newblock Population requiring preventive chemotherapy for schistosomiasis
  (data adapted from the {WHO} global health observatory), 2026{\natexlab{b}}.
\newblock URL
  \url{https://ourworldindata.org/grapher/population-requiring-preventive-chemotherapy-for-schistosomiasis}.
\newblock Data downloaded 13 November 2026.

\bibitem[Pedregosa et~al.(2011)Pedregosa, Varoquaux, Gramfort,
  et~al.]{pedregosa2011}
Fabian Pedregosa, Ga{\"e}l Varoquaux, Alexandre Gramfort, et~al.
\newblock Scikit-learn: machine learning in {Python}.
\newblock \emph{Journal of Machine Learning Research}, 12:\penalty0 2825--2830,
  2011.
\newblock URL \url{https://jmlr.org/papers/v12/pedregosa11a.html}.

\bibitem[Peel and McLachlan(2000)]{peel2000}
David Peel and Geoffrey~J. McLachlan.
\newblock Robust mixture modelling using the $t$ distribution.
\newblock \emph{Statistics and Computing}, 10\penalty0 (4):\penalty0 339--348,
  2000.
\newblock URL \url{https://doi.org/10.1023/A:1008981510081}.

\bibitem[Robins et~al.(1994)Robins, Rotnitzky, and Zhao]{robins1994}
James~M. Robins, Andrea Rotnitzky, and Lue~Ping Zhao.
\newblock Estimation of regression coefficients when some regressors are not
  always observed.
\newblock \emph{Journal of the American Statistical Association}, 89\penalty0
  (427):\penalty0 846--866, 1994.
\newblock URL \url{https://doi.org/10.1080/01621459.1994.10476818}.

\bibitem[Rosenbaum and Rubin(1983)]{rosenbaum1983}
Paul~R. Rosenbaum and Donald~B. Rubin.
\newblock The central role of the propensity score in observational studies for
  causal effects.
\newblock \emph{Biometrika}, 70\penalty0 (1):\penalty0 41--55, 1983.
\newblock URL \url{https://doi.org/10.1093/biomet/70.1.41}.

\bibitem[Rubin(1974)]{rubin1974}
Donald~B. Rubin.
\newblock Estimating causal effects of treatments in randomized and
  nonrandomized studies.
\newblock \emph{Journal of Educational Psychology}, 66\penalty0 (5):\penalty0
  688--701, 1974.
\newblock URL \url{https://doi.org/10.1037/h0037350}.

\bibitem[Tsiatis(2006)]{tsiatis2006}
Anastasios~A. Tsiatis.
\newblock \emph{Semiparametric Theory and Missing Data}.
\newblock Springer, New York, 2006.

\bibitem[Virtanen et~al.(2020)Virtanen, Gommers, Oliphant,
  et~al.]{virtanen2020}
Pauli Virtanen, Ralf Gommers, Travis~E. Oliphant, et~al.
\newblock {SciPy} 1.0: fundamental algorithms for scientific computing in
  {Python}.
\newblock \emph{Nature Methods}, 17:\penalty0 261--272, 2020.
\newblock URL \url{https://doi.org/10.1038/s41592-019-0686-2}.

\bibitem[{WHO Regional Office for Africa}(2026)]{espen}
{WHO Regional Office for Africa}.
\newblock {ESPEN} data portal: schistosomiasis, implementation-unit data for
  the democratic republic of the congo, 2014--2025, 2026.
\newblock URL \url{https://espen.afro.who.int/}.
\newblock Expanded Special Project for Elimination of Neglected Tropical
  Diseases; data downloaded 13 November 2026.

\bibitem[{World Health Organization}(2022)]{who2022}
{World Health Organization}.
\newblock \emph{{WHO} Guideline on Control and Elimination of Human
  Schistosomiasis}.
\newblock World Health Organization, Geneva, 2022.
\newblock URL \url{https://www.who.int/publications/i/item/9789240041608}.
\newblock ISBN 978-92-4-004160-8.

\bibitem[{World Health Organization}(2026)]{who_pct}
{World Health Organization}.
\newblock Preventive chemotherapy data portal: Schistosomiasis, country-level
  data 2006--2025, 2026.
\newblock URL
  \url{https://www.who.int/teams/control-of-neglected-tropical-diseases/data-platforms/pct-databank/schistosomiasis}.
\newblock Data downloaded 13 November 2026.

\bibitem[Worrell and Mathieu(2012)]{worrell2012}
Caitlin Worrell and Els Mathieu.
\newblock Drug coverage surveys for neglected tropical diseases: 10 years of
  field experience.
\newblock \emph{The American Journal of Tropical Medicine and Hygiene},
  87\penalty0 (2):\penalty0 216--222, 2012.
\newblock URL \url{https://doi.org/10.4269/ajtmh.2012.12-0167}.

\end{thebibliography}
\end{document}